\documentclass[conference,compsoc]{IEEEtran}

\ifCLASSOPTIONcompsoc
  \usepackage[nocompress]{cite}
\else
  \usepackage{cite}
\fi

\ifCLASSINFOpdf
\else
\fi
\usepackage{glossaries}
\usepackage{hyperref}
\usepackage{multirow}
\usepackage{graphicx}
\usepackage{subcaption}
\usepackage{pdflscape}
\usepackage{enumitem}

\usepackage{rotating}

\begin{document}
\newacronym{WHO}{WHO}{World Health Organization}
\newacronym{GDP}{GDP}{Gross Domestic Product}
\newacronym{IOT}{IoT}{Internet of Things}
\newacronym{HITLCPS}{HITLCPS}{Human-in-the-Loop Cyber-Physical-System}
\newacronym{ISABELA}{ISABELA}{Iot Student Advisor and BEst Lifestyle Analyser}
\newacronym{GE}{GE}{Generic Enablers}
\newacronym{API}{API}{Application Program Interfaces}
\newacronym{BLE}{BLE}{Bluetooth Low Energy}
\newacronym{SDK}{SDK}{Software Development Kit}
\newacronym{SAM}{SAM}{Self-Assessment Manikin}
\newacronym{DGS}{DGS}{Portuguese National Health System}
\newacronym{MHApp}{MHApp}{Mental Health Application}
\newacronym{ICT}{ICT}{Information and Communication Technologies}

%
\title{Social Sensing and Human in the Loop Profiling during Pandemics: the Vitoria application }


\author{\IEEEauthorblockN{J. Fernandes}
\IEEEauthorblockA{Center of Informatics \\
and Systems $|$ CISUC,\\
University of Coimbra,\\
Coimbra, Portugal\\
jmfernandes@dei.uc.pt}
\and
\IEEEauthorblockN{J. Sá Silva}
\IEEEauthorblockA{Institute for Systems\\ 
Engineering and Computers\\
at Coimbra $|$ INESC,\\
University of Coimbra,\\
Coimbra, Portugal
\\sasilva@deec.uc.pt}
\and
\IEEEauthorblockN{A. Rodrigues}
\IEEEauthorblockA{Coimbra Business School
\\Research Centre $|$ ISCAC,\\
Polytechnic of Coimbra and\\
Center of Informatics \\
and Systems $|$ CISUC,\\
University of Coimbra,\\
Coimbra, Portugal\\andre@iscac.pt}
\and

\IEEEauthorblockN{F. Boavida}
\IEEEauthorblockA{Center of Informatics \\
and Systems $|$ CISUC,\\
University of Coimbra,\\
Coimbra, Portugal\\
boavida@dei.uc.pt}
\and
\IEEEauthorblockN{R. Gaspar}
\IEEEauthorblockA{Catolica Research Centre for \\Psychological Family and\\Social Wellbeing,\\
Catholic University\\
Lisbon, Portugal\\rgaspar@ucp.pt}
\and
\IEEEauthorblockN{C. Godinho}
\IEEEauthorblockA{Catolica Research Centre for \\Psychological Family and\\Social Wellbeing,\\
Catholic University\\
Lisbon, Portugal\\cristinagodinho@ucp.pt}
\and
\IEEEauthorblockN{R. Francisco}
\IEEEauthorblockA{Catolica Research Centre for \\Psychological Family and\\Social Wellbeing,\\
Catholic University\\
Lisbon, Portugal\\ritafrancisco@ucp.pt}

}


%


\maketitle

\begin{abstract}

As the number of smart devices that surround us increases, so do the opportunities to leverage them to create socially- and context-aware systems. Smart devices can be used for better understanding human behaviour and its societal implications. As an example of a scenario in which the role of socially aware systems is crucial, consider the SARS-CoV-2 pandemic. In this paper we present an innovative Human-in-The-Loop Cyber Physical system that can collect passive data from people, such as physical activity, sleep information, and discrete location, as well as collect self-reported data, and provide individualised user feedback. In this paper, we also present a three and a half months field trial implemented in Portugal. This trial was part of a larger scope project that was supported by the Portuguese National Health System, to evaluate the indicators and effects of the pandemic. Results concerning various applications usage statistics are presented, comparing the most used applications, their objective and their usage pattern in work/non-work periods. Additionally, the time-lagged cross correlation between some of the collected metrics, Covid events, and media news, are explored. This type of applications can be used not only in the context of Covid but also in future pandemics, to assist individuals in self-regulation of their contagion risk, based on personalized information, while also function as a means for raising self-awareness of risks related to psychological wellbeing.

\end{abstract}


%
\IEEEpeerreviewmaketitle

\section{Introduction}

The \gls{WHO} first declared COVID-19 a world health emergency in January 2020. Since 
the virus was first diagnosed in Wuhan, China, it has been detected in virtually every country. The number of cases of COVID-19 infection surpassed 250 million, being responsible for more than 5 million deaths \cite{COVIDLiv50Online}.

Furthermore, apart from the tragic cost in human lives, the pandemic has affected the world’s economy. Most countries went through long periods of lockdown, with considerable impact on the social system.  In \cite{fernandes2020economic}, a study made with the data from 30 countries has shown that, on average, lockdowns had a negative effect of 2.8\% on the \gls{GDP} of these countries. The authors also predicted that the global \gls{GDP} can decrease 2.5-3\%, for each additional lockdown month. In \cite{weiss2020global}, the authors have also verified the existence of this negative effect, showing that in America the pandemic lead to a decrease of 4.8\% in the \gls{GDP}, and contributed to unemployment insurance claims for more than 36.5 million Americans.
 
Apart from the human and economic cost, other authors have also highlighted the psycho-social effects of the COVID-19 pandemic \cite{haleem2020effects}. These are caused by the interference, restrictions, and changes in daily routines caused by the pandemic. Some of these changes include: the lack of face-to-face services, cancellation or postponement of large-scale sports and tournaments, travelling restrictions, disruption of celebration of cultural, religious and festive events, social distancing from peers and family members, closure of recreation places (e.g., hotels, restaurants, theatres, cinema, bars, etc.), postponement of examinations, etc. These lead to an increase in general stress among the population \cite{santomauro2021global}. In Italy a study was conducted with adolescents, to evaluate how the pandemic affected their lifestyle and emotional state \cite{buzzi2020psycho}.
The results pointed to the resilience of the younger generations when faced with uncertain and unpredictable situations. However, other studies have pointed out the negative effects of the restrictions caused by the ongoing pandemic \cite{afonso2020impact}.

Although, some studies have sought to evaluate the impact of the ongoing pandemic, having human-beings as the subject of research is always challenging. This is particularly true when performing longitudinal studies, that is, studies that collect several samples of data from the same population and study the changes over a period of time. However, with the proliferation of \gls{IOT} devices, it is now possible to unobtrusively collect large amounts of data in order to better understand humans and their behaviours. Additionally, these devices are equipped with sensors that can retrieve data from their surroundings, providing a glimpse of their surrounding context. Thus, if properly configured, and provided privacy concerns are strictly met, \gls{IOT} devices are able to obtain data about human cognition, emotions and behaviours, as well as the environment that surrounds them. By doing so, we can move to the concept of the \gls{HITLCPS} where, in addition to gathering and processing data about humans and their surroundings, the system actively provides user feedback \cite{nunes2018practical}.

Although there was an effort to create contact tracing applications across several countries in order to avoid community transmission of the SARS-CoV-2 virus \cite{guillon2020attitudes}, these applications suffered from a lack of adoption from end users, mainly due to cybersecurity and privacy concerns and a general lack of trust in the governments \cite{altmann2020acceptability}. Additionally, these applications did not offer a broad understanding of human behaviour and associated changes due to the pandemic. As such, applications that can monitor people’s behaviours, lifestyles, activities and their psychological wellbeing, and that can be used to provide meaningful advantages to their users, are still needed.

In this paper we present the Vitoria HITLCPS mobile application, as well as a field trial in the context of the Covid-19 pandemic. This application is able to track several aspects of human behaviour, such as monitoring user discrete location, physical activity levels, mental state, number of surrounding people, used means of transportation, among several other aspects. The acquired data is explained in detail in sections \ref{sec:data_aquisition} and \ref{sec:active_data}. Furthermore, Vitoria comprises an individualized feedback system designed to offer user guidance, concerning physical and emotional well-being, as well as covid-related information. Since we are dealing with personal data, privacy is a must. As such, all collected data is duly anonymized, and no personal identifiers are collected from the user. Furthermore, all communications and data storage are secured.

A three and a half months trial was conducted in Portugal, from the 1$^{st}$ of February 2021 to the 13$^{th}$ of May 2021, with 19 participants who used the mobile application. During this trial, several Covid-related events occurred, which included various deconfinement periods. As such, this longitudinal study, offers an important outlook on the participants’ well-being and daily lives during different restrictive periods. From this study several preliminary results are presented, such as a summary of the most used applications and their objective, as well as the correlation between the collected data and the metrics commonly used by the media to communicate the pandemic situation. The contributions of this paper can be summarized as follows:

\begin{itemize}
    \item A mobile \gls{HITLCPS} application, called Vitoria, was developed, to assist people in the context of SarsCov-2 pandemic and future pandemics. This application collects behavioural data as well as environment data, with the objective of providing individualised feedback and guidance, through a chatbot. 
    
    \item A trial performed with the Vitoria application to collect data during three and a half months in Portugal, during the covid-19 pandemic, with 19 subjects. This trial was developed in the scope of a bigger project which included the  \gls{DGS} as a partner. This longitudinal study offers an important outlook on the participants well-being and daily lives during different restrictive periods.
    
    \item Several preliminary results are presented, including a summary of the most used end-user applications, as well as the correlation between the collected data and the metrics commonly used by the media to communicate the pandemic situation.
    
\end{itemize}

The rest of the paper is structured as follows. In section \ref{sec:related_work}, we present an overview of related work and explain the differences and added value of Vitoria. Section \ref{sec:ss_framework} provides information on the ResiliScence 4 COVID-19 project, in the scope of which the Vitoria application was developed, along with an overview of other methods, collected data and some of their results. Section \ref{sec:system_implementation} is dedicated to the detailed presentation of the Vitoria application. We present the architecture and describe the implementation of the system. The section also includes information on the acquisition of data, its processing, and privacy measures. Section \ref{sec:trials} presents the field trial and respective results. The conclusions and guidelines for further work are presented in Section \ref{sec:conclusions}.

\section{Related work}
\label{sec:related_work}

The main goal of the Vitoria system is to monitor the day to day emotional and physical states of users and their surrounding context, in order to evaluate possible behavioural patterns of contagion risk exposure and relevant wellbeing indicators, and to provide personalised user feedback. Understanding how people are affected and behave during pandemics is important for the development of more efficient and personalized plans and measures, for tackling future Covid-19 waves, or even future pandemics. In this section we will cover related state-of-the-art work that focused on aspects of understanding human behaviour as well as the effects of the ongoing pandemic on people’s wellbeing. In addition, we will also cover contact tracing applications, as these mobile applications were one of the most used approaches to deal with the Covid-19 pandemic.

The Covid-19 pandemic required social distancing and, in some periods, lockdown measures. This, in turn, led to several changes in our daily routines. Lockdowns imposed a shift in school \textit{”normality”}, making most children change to a paradigm of virtual classes. In \cite{limone2021psychological}, the authors evaluated the effects that this shift had in children, particularly the effects of the increase in digital technology use. In this work the authors highlighted that the increase in the usage of technology during the pandemic had both positive and negative effects. Digital technologies allow us to bypass the separation created by the social distance, allowing us to connect with others. However, they may also be a source of risks related to psychological wellbeing, and contribute to depression, anxiety, sleep problems, irritability, and cognitive impairment. In \cite{francisco2020psychological}, the authors also found significant changes in lifestyle habits of children and adolescents (e.g., increased screen-time, reduced physical activity, more sleep hours per night), associated with an increase in psychological and behavioural symptoms, such as anxiety, mood, food intake, cognitive and behavioural changes.

In \cite{talevi2020mental}, the authors survey a series of works that evaluated the mental health of people during the pandemic. They concluded that the ongoing pandemic has had a huge psychological impact on individuals, from common people to people that were infected, and people previously diagnosed with mental health issues, and even healthcare workers. In \cite{greenberg2020mental}, it was also explored how healthcare workers were negatively affected by the ongoing pandemic, due to working extremely long hours in high-stress environments, or even the exposure to trauma and being faced with moral dilemmas when trying to deliver high-quality care.

Additionally, some studies evaluated the influence of the Covid-19 pandemic in physical activity levels \cite{caputo2020studies}, with evidence pointing towards a decrease in physical activity levels, due to home confinement and social distancing measures. Furthermore, the positive correlation between physical activity and mental health was also pointed out.

Other studies have also used questionnaires to evaluate the perception of threat of the Covid-19 pandemic \cite{perez2020questionnaire}. For instance, they showed that it was possible to assess the general perception of the Spanish population using a short questionnaire. However, this type of study can only evaluate the momentary perception, as the use of longer questionnaires constrains the possibility of doing a longitudinal study.

Mobile phones and wearables have also been leveraged to respond to the public health crisis imposed by the COVID-19 pandemic. A systematic review of health apps for COVID-19 \cite{almalki2021health} has identified five main purposes of the identified apps. The most common were apps developed by governments or national health authorities to encourage users to track their personal health, including self-assessment for the identification of possible infection, symptoms’ monitoring, mood tracker, medication trackers, diagnosis recorders used during self-quarantine, etc. The second most common purpose was to raise awareness about COVID-19, including providing basic health information and advice, presenting statistics, latest news and updates about the pandemic, providing information or interactive maps of active cases and medical facilities, etc. Less common purposes were the management of COVID-19 exposure risk, providing health monitoring by healthcare providers (such as making medical appointments, virtual medical consultations, remote monitoring, helpline, etc.) and conducting research studies.

When considering mobile phones and their use in relation to the management of COVID-19 exposure risk, the most common line of work relates to contact tracing applications. The main goal of these applications is trying to limit the propagation of the SARS-CoV-2 virus, by detecting in a short period of time the past contacts of a person that tested positive for the virus. By knowing the past contacts of a positively diagnosed person, it is possible to enforce preventive measures such as, for instance, putting those contacts in a preventive quarantine to mitigate the spread of the virus. With the outburst of the virus, there was an emergence of several of these apps, in part because most countries tried to implement their own contact tracing mobile application. In Portugal, where our trials occurred, the government took this approach as well, with the creation of the "STAYAWAY Covid" app \cite{STAYAWAY38:online}. This type of application normally uses the smartphone's Bluetooth  capabilities to transmit and receive individual hashes that can then be used to identify the contacts of each person and notifying users who were within proximity of someone who tested positive for the virus.

However, these applications are merely intended to detect possible past contacts, and are not able to offer more information about their users, nor to provide feedback to users on relevant information in relation to their risk of exposure to COVID-19, such as, for example, the number of contacts they had or the time they spent out of their home, which have been proved to be one of the most relevant features of human behaviour self-regulation \cite{carver1981self}. Tailored feedback represents the most individualised type of feedback, and research has pointed to its potential for increasing the effectiveness of behaviour change interventions, over providing generic or targeted information \cite{diclemente2001role}. However, tailored feedback requires personal information provided by some kind of assessment procedure, which can be burdensome for the user if it is only acquired by active data acquisition procedures, requiring, for example, the user to fill in some measurements. In this regard, mobile phones provide an effective way of unobtrusively sensing human behaviour and delivering tailored feedback.

As stated before, the Vitoria application monitors the day to day emotional and physical states of users and their surrounding context. The objective is to detect and identify possible profiles and behaviours that are common in this and possibly in future pandemics, with a view to providing tailored feedback to users that is vital for them to self-regulate their behaviour, in order to better manage their risk of exposure to SARS-CoV-2 contagion. In what concerns privacy, Vitoria differs from the aforementioned applications, as we do not intend and do not need to identify specific people. As such, all the collected data is anonymized, making it impossible for our system or for third-party systems to identify the users. Moreover, as far as we know, our system offers the most complete outlook on the user’s daily lives, and is the only application that closes the loop, by incorporating a personalised feedback mechanism.

In recent years, several applications have been built and studied to help promote positive mental health and wellbeing in the general population, that we will collectively refer to as \gls{MHApp}. Despite promising results pointing to the reduction of mental health problems and/or to promoting well-being or emotion regulation, the lack of generality of the findings and the low adherence to \gls{MHApp} usage are commonly reported as limitations \cite{eisenstadt2021mobile}. On other hand, several \gls{MHApp}s were also developed specifically during the COVID-19 pandemic to help reduce mental health problems in the general population \cite{jaworski2021exploring} and in healthcare professionals \cite{fiol2021mobile}, but also with reduced engagement.

Additionally, there are other pieces of work that use mobile applications or \gls{IOT} devices to monitor people, some of them being identified in \cite{kondylakis2020covid}. However, those studies focus mainly on contact tracing or data sharing from patient to healthcare professionals. As far as we known, the Vitoria system is the only system that leverages smartphones and smartwatches to obtain a comprehensive overview of the users psychological and contextual state, as well as give them tailored feedback.

\section{ResiliScence - A social sensing framework grounded on the layering method}
\label{sec:ss_framework}

In any pandemic, studying the virus’ transmission channels and dissemination is central to controlling it and mitigating its impact on the various elements of the social system and individuals’ health. Also, a central factor in the evolution of the pandemic is human behaviour, which largely determines both the spread of viruses and their control. Hence, public health strategies should motivate the public’s adherence to the different recommendations for contagion prevention behaviours. To increase adherence, it is important to understand the factors that can predict behaviour and longitudinally monitor changes on how people evaluate and respond to the unfolding events during a pandemic. This was the goal of the project “ResiliScence 4 COVID-19 - Social Sensing \& Intelligence for Forecasting Human Response in Future COVID-19 Scenarios, towards Social Systems Resilience” \footnote{Project n. 439 – Research 4 COVID-19 - 2ª Edition, Portuguese Foundation for Science and Technology; Gaspar, Rodrigues, Raposo, et al., 2021}. This project was coordinated by the Católica Research Centre for Psychological, Family and Social Wellbeing (CRC-W) – Catholic University of Portugal, with the following partners: University of Coimbra, Faculty of Sciences and Technology; Portuguese Directorate-General for Health; National Health Institute Dr. Ricardo Jorge; ISPA-Instituto Universitario; Portuguese Order of Psychologists.

This project proposed a social sensors framework – the ResiliScence Approach \cite{gaspar2021striving} – aimed at monitoring social perceptions of systemic risks (derived by evaluations of the perceived demands posed by the pandemic and the perceived resources to cope with these), self-reported protective behaviours against SARS-CoV-2 contagion and their predictors, and subjective and objective indicators of exposure to risks associated with SARS-CoV-2 contagion. This framework derives from the concept of social sensing as a \textit{“form of crowdsourcing that involves systematic analysis of digital communications to detect real-world events”} \cite{arthur2018social}. This concept was extended in the proposed framework to include not only digital communications (e.g., mediated by social media) but, more broadly, \gls{ICT}. This is because \gls{ICT} can mediate individuals’ interactions with the surrounding social and physical environment through smartphone and smartwatch-based applications (as shown in the current article, through the Vitoria application). Also relevant in this regard is self-reported data collected through online data collection platforms (Qualtrics™), as these can also function as platforms for social sensing. This inclusion in addition to digital communications, enables a social sensing framework through the “layering method” \cite{barnett2003social,domingos2022pandemia}.

\begin{figure*}[!t]
\centering
\includegraphics[width=7in]{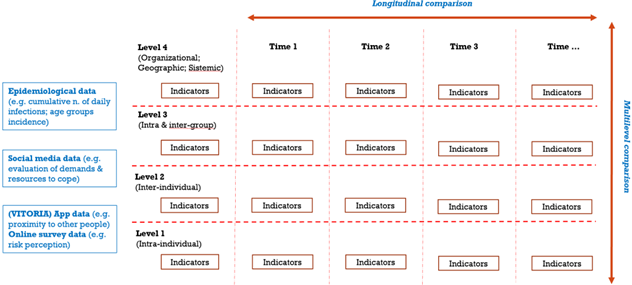}
\caption{Graphical representation of a social sensing framework through the layering method with examples of data collection.}
\label{fig:framework_ss}
\end{figure*}

For an effective crisis detection and monitoring mechanism to be implemented, a “layering method” such as the one proposed in \cite{barnett2003social} should be applied. This method assumes that different “layers” of data can be collected and interpreted, co-occurring, within each time period and between periods. For this method to be applied, there must be at least two “data layers”, i.e., collected at least in two levels of analysis (e.g., individual and group). As proposed in \cite{domingos2022pandemia}, a social sensing layering method should have the following characteristics:

\begin{enumerate}[leftmargin=*]

\item Data must be at different levels of analysis – for example, data may be collected at an intra- and inter-individual level (e.g., cognition, emotions, behaviours of each individual and/or differences between individuals), at an intra- and inter-group level (e.g., social representations of danger and risk, within a population group, community and/or between groups and communities), at an intra and inter-organizational level (e.g., formats, contents and uncertainty communication channels) and at macro geographically defined levels (e.g., counties, regions, countries), and/or that make up all the mentioned levels, corresponding to the social system (e.g., crisis model as evaluated by the social system).

\item The time dimension should be used as a systematic focus of analysis – whenever possible data should be available over a period of time, and different levels of data should correspond to the same period of time. The period or periods of time (e.g., time during which the hazard is present, from its emergence to its reduction/elimination) should previously be established, thus allowing data to be compared at various levels.

\item The analysis should make it possible to evaluate relationships between variables/factors at a given moment and/or over time – the same indicators should be evaluated between levels in a given period of time, but also, ideally, the comparison of indicators between periods of time, in the different levels. In this way, the focus of the analysis will be the “overlap” or co-occurrence of changes at different levels over time.
\end{enumerate}

The characteristics of the application of a social sensing layering method can be graphically represented as a multilevel longitudinal monitoring system, as represented in Figure \ref{fig:framework_ss}, with examples of the type of data that can be collected at the “frontier” between levels.

 The social media data monitoring in co-occurrence with the pandemic related events (i.e., registered changes in the epidemiological situation) was already described elsewhere (\cite{gaspar2021striving}, \cite{gaspar2021resiliscence} and \cite{domingos2022pandemia}) and, thus, will not be described here. Other examples with other types of data can also be found in \cite{galesic2021human}.

Concerning the self-reported behaviour monitoring system aimed at preventing the risk of contagion by SARSCoV-2, implemented through a longitudinal web-based survey with data collected through the Qualtrics™ online platform, a short summary is presented here. This system assumed that to increase the likelihood of performing protective behaviours, it is important to foster the internal conditions (characteristics of each individual) and external conditions (characteristics of the physical-social context in which they are immersed) that facilitate this occurrence. The COM-B Model, developed in \cite{michie2011behaviour} and \cite{michie2013behavior}, considers the physical and psychological Capabilities (C) of each individual, the physical and social Opportunities (O) to perform the behaviours, and the reflexive (deliberate, conscious) and automatic (spontaneous, not conscious) Motivations (M), as well as the interrelationships between these variables, as the determinants of Behaviour (B). In other words, to implement protective behaviours against SARSCoV-2 contagion, each individual needs to have the essential capabilities, opportunities, and motivations to do so.

Based on the COM-B Model, a longitudinal web-based survey was carried out during the first year of the pandemic in Portugal, to assess a set of variables representing each category of determinants (COM), as well as the frequency of behaviours (B) reported by the study participants. This was done in three waves - August, October, and November 2020 - with a representative sample of the Portuguese population in terms of its distribution by age, gender and region of the country. A total of 333 people recruited from the Online Study Panel (PES) of the Catholic University of Portugal participated in the three waves. All of them gave their informed consent to participate in the online survey through the Qualtrics™ platform. Participants were between 18 and 59 years old, with a mean age of 38.49 years. Of these,
169 (50.8\%) were women. As for the region of residence,
114 lived in the North region of Portugal (34.2\%), 73 in the Central Region (21.9\%), 85 in the Lisbon Metropolitan Area (AML; 25.5\%), 37 in the South Region (11.1\%), and 24 in the Azores and Madeira regions (7.21\%).

Psychometrically validated measurement scales were used, or adapted by the project team (e.g., scale of protective behaviours against SARS-CoV-2 contagion) based on existing scales in the literature (information on the scales used can be found in the project’s final scientific report \cite{gaspar2021resiliscence}. The scales sought to measure protective behaviours against SARS-CoV-2, capabilities (e.g., health literacy), motivations (e.g., behavioural intention) and opportunities (e.g., social norms) as determinants of these behaviours. The eight global protective behaviours analysed were: 1) mask use; 2) hands hygiene; 3) respiratory etiquette; 4) physical distancing; 5) ventilation of spaces; 6) reduction of contacts; 7) cleaning of surfaces; and 8) contact avoidance associated with symptoms self-vigilance (5-point scale: 1 – Never; 2 – Almost never; 3 – Sometimes; 4 – Often; 5 – Whenever possible).

Mask use and hand hygiene were the most reported protective behaviours globally. Mask use increased significantly between August (Mean = 4.05) and November (Mean = 4.63), p $<$ .001, but not between October (Mean = 4.23) and November. The frequency of hand hygiene also increased between August (Mean = 4.01) and October (Mean = 4.51), p $<$ .001, and between August and November (Mean = 4.55), p $<$ .001. Regarding physical distancing, there was a marginal increase between August (Average = 4.38) and November (Average = 4.51), p $<$ .10, and a significant increase between October (Average = 4.37) and November, p $<$ .05. Regarding the ventilation of spaces, there were significant increases between all periods considered: August (Average = 3.30), October (Average = 3.70) and November (Average = 4.42), p $<$ .001. Regarding the reduction of contacts, there was a significant increase between October (Average = 4.22) and November (Average = 4.40), p $<$ .01. In terms of respiratory etiquette, avoidance of contacts associated with self-monitoring of symptoms, and cleaning of surfaces, there were no significant differences between the different periods. In general, it can be concluded that almost all behaviours had a higher report in November than in previous periods.

Concerning the predictors of protective behaviours and particularly focusing on the last wave data, the intention to perform each protective behaviour was a significant predictor of most behaviours –except respiratory etiquette and surface cleaning. Risk perception, social norm and health literacy did not prove to be predictors of different prevention behaviours. Sociodemographic variables and variables related to the risk of contagion by SARS-CoV-2 were only significant in predicting adherence to prevention behaviours, for physical distancing, contact avoidance associated with symptom surveillance, and respiratory etiquette, for which the women reported greater adherence. More detailed results can be found in the project report \cite{gaspar2021resiliscence}.

Given this example of self-reported behaviour monitoring through a web-based longitudinal survey, it is clear that (perceived) behavioural changes have occurred across time. However, there were no statistically significant changes in risk perception reported across the 3 waves of the study, namely between August (Average = 3.59), October (Average = 3.72), and November (Average = 3.80), p $>$ .05 (considering a 5-point scale). Despite the common-sense view at a societal level (e.g., policy makers; health authorities) that risk perception is a main driver of protection behaviours, these results raise the question whether indeed risk perception can account for changes in self-reported behaviours. Hence, it is important to “add” more layers to the analysis, given that longitudinal surveys provide an incomplete view of the drivers of human behaviour. For this reason, it is also relevant to implement a monitoring system that targets indicators of exposure to the risk of contagion by SARS-CoV-2, based on social sensors. Next an example of this is presented, based on the developed smartphone-based Vitoria application.

\section{The Vitoria System}
\label{sec:system_implementation}

The Vitoria system is based in our previous work on the \gls{ISABELA} case study \cite{fernandes2020isabela}. Our previous case study aimed to monitor and profile students, as well as improve their academic outcomes. However, in the Vitoria case study we aim to monitor and profile the general individual, and to determine how his/her behaviour is affected by changes in the pandemic situation, through the layering method presented in the previous section. Understanding these changes can lead to the implementation of more effective measures to fight the pandemic. Vitoria directly translates to \textit{”Win”} in Portuguese, and with this application we aim to create a tool that would help to understand and fight the progression of the current and future pandemics.

\begin{figure}[!t]
\centering
\includegraphics[width=3in]{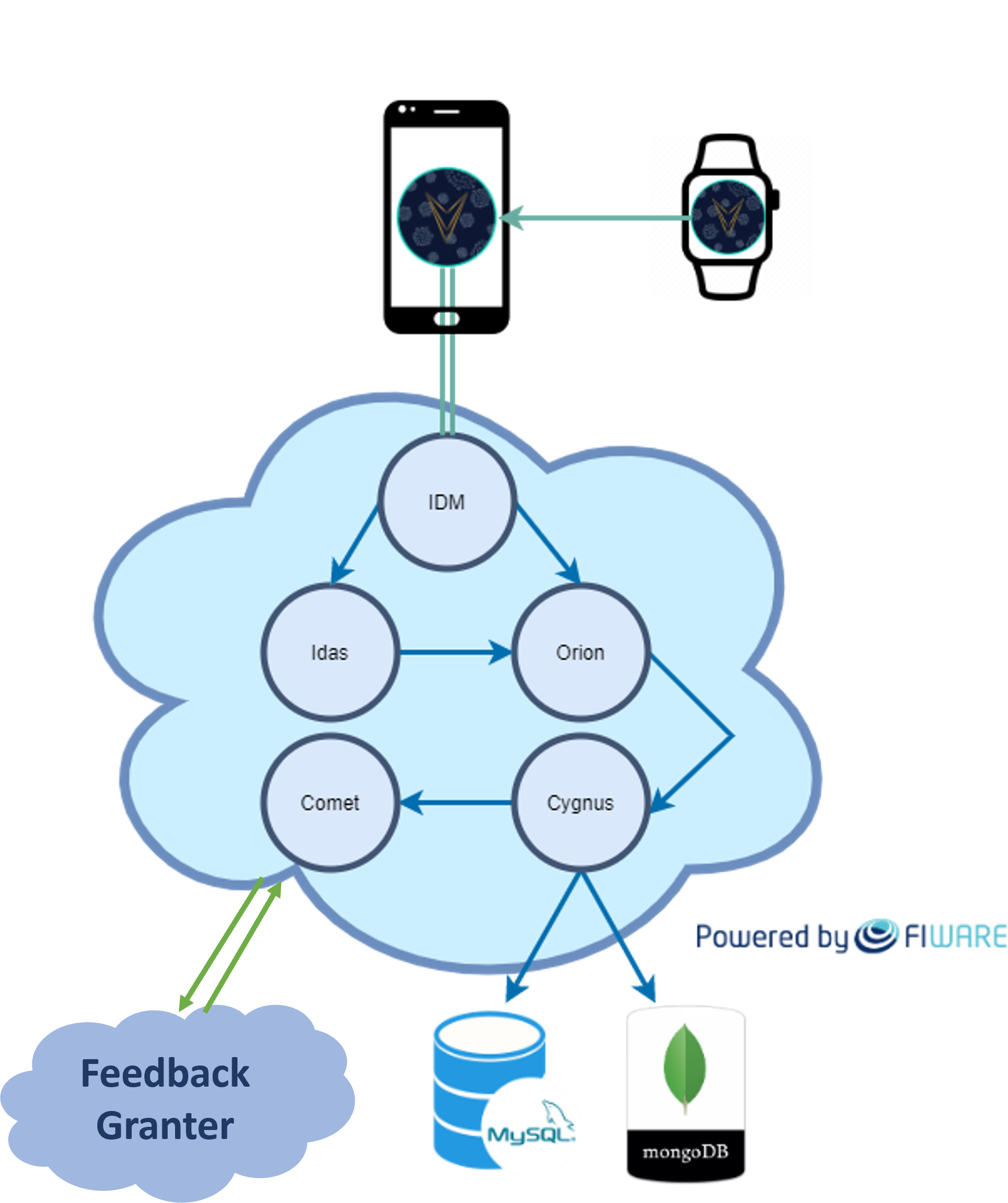}
\caption{Vitoria system Architecture.}
\label{fig:architecture}
\end{figure}

Although, \gls{ISABELA} and Vitoria target different use cases, they share many architectural similarities, such as the FIWARE-based backend \cite{FIWAREOp8:online} and the Android-based Mobile and Wear applications. As can be seen in Figure \ref{fig:architecture}, the system is made up of a smartwatch and smartphone applications that deal with data acquisition and users feedback, and the backend that is able to manage and persistently store the collected data. In this section we present implementation details of the Vitoria system.

\subsection{Backend Architecture}
\label{sec:backend_achitecture}

As previously stated, the Vitoria system was built on the FIWARE ecosystem \cite{FIWAREOp8:online}. The FIWARE project offers a set of open-source components called \gls{GE} to deal with some of the challenges of \gls{IOT}-enabled systems, and to accelerate development.

The \gls{GE}s were designed to target specific problems in the \gls{IOT} domain, such as data heterogeneity, protocol heterogeneity, data subscriptions, or authentication and security. \gls{GE}s also offer several \gls{API}s that help third party applications to communicate with them. In the Vitoria system, we used several \gls{GE}s from the FIWARE project, namely IDAS, ORION, IDM, Cygnus, and Comet, as can be seen in Figure \ref{fig:architecture}.

The IDAS \gls{GE} is able to translate several common \gls{IOT} communication protocols to the FIWARE-enabled HTTP/NGSI10 protocol \cite{idas:online}. Although in the current state the system only communicates with the users’ smartphones, the use of this \gls{GE} allows us to add new \gls{IOT} devices to the system without making any changes. In the future we foresee the use of \gls{IOT} boxes similar to the ones deployed in \cite{fernandes2020isabela}, to not only gather user data but also data on surrounding environments (e.g., temperature, humidity, sound).

In addition to adaptability in terms of communication protocols, a \gls{HITLCPS} system needs to be scalable and support ever-changing dynamic models. The FIWARE project adopts the NGSI10 information model, previously developed by OMA \cite{krvco2014designing}. This model uses the JSON format, and is based on entities, which have a set of attributes and its own type. To manage these entities, our system uses the ORION \gls{GE} \cite{orion:online}. Using the \gls{API} from this \gls{GE}, it is possible to create/delete/retrieve entities or update existing ones. This \gls{GE} allows external systems to make data subscriptions to entities with specific type and attribute rules.

Due to its context management nature, the ORION \gls{GE} is not made with the purpose of storing historic data. For this purpose, the FIWARE ecosystem uses the Cygnus \gls{GE} \cite{cygnus:online} as a connector, which is capable of persistent data storage. This module works by using the Apache Flume technology and by subscribing to the ORION entity changes. Thus, when a new entity is created on ORION, or when an existing one is updated, a notification is sent to the Cygnus \gls{GE}. The latter module will then put the information contained in the notification into a specific channel and forward it to a third-party data storage (i.e., MySQL \cite{MySQL29:online}, MongoDB \cite{MongoDBt67:online}).

Databases normally serve the sole purpose of storing data and, as such, they do not provide \gls{API}s to allow applications to retrieve the data. Hence, we need a different module that can retrieve the historic stored data. The FIWARE platform catalogue has a \gls{GE} that handles this issue, namely the Short-Term History or STH-Comet \gls{GE} \cite{sthcomet:online}. This \gls{GE} provides a RESTful \gls{API} which allows external systems to perform historic and aggregation queries. This \gls{API} is used by the Vitoria smartphone application and the other backend modules to retrieve data.

Additionally, security and privacy are some of the most important requirements of a \gls{HITLCPS} system. In order to protect communications between \gls{GE}s, end user devices, and applications, the Identity Management \gls{GE} called Keyrock/IDM \cite{idm:online}, was used. This \gls{GE} allows the implementation of device, user, and application authentication and security, as well as authorization policies. Furthermore, in order to meet all requirements from EU privacy laws, such as the General Data Protection Regulation of May 2018 \cite{hoofnagle2019european}, all data is anonymised.

In addition to the FIWARE \gls{GE}s, a custom microservice, called Feedback Granter, was created for the Vitoria system. As can be seen in Figure \ref{fig:architecture}, this module is able to communicate with the STH-Comet \gls{GE} to retrieve historic user data. This module is responsible for distributing the users by two feedback groups. Furthermore, this module is also able to generate the feedback metrics that are shown to the user, based on the raw data stored in the databases. The Feedback system is explored in more detail in section \ref{sec:feedback}.

\subsection{Passive Data Acquisition}
\label{sec:data_aquisition}

Both the mobile application and the smartwatch application from the Vitoria system serve primarily as means to obtain data from the user. Most of the metrics collected come from the smartphone, while the smartwatch is primarily used to acquire health-related data (i.e., step count, activity, and heart rate).

One of the most important metrics, when considering the pandemic context is user location. Due to privacy concerns, our system does not store any GPS information. Instead, the system is able to detect the discrete location between two available options, \textit{home} or \textit{other}. In order to infer if the user is at home or in other places, when the users first configure their app, they are prompted to scan the available networks and select the name of their home network. The name of their home network is then stored in a local database on the smartphone. Due to privacy concerns this information never leaves the smartphone. The application is then able to periodically scan the networks and compare the scan results with the stored information, in order to infer if the user is at \textit{home} or not.

Apart from mandatory confinement periods, during the pandemic in Portugal there were additional traveling restrictions between municipalities in certain periods of time. In addition, the analysis of the number of cases was carried out at municipality level. As such, user movements between municipalities were also an important metric to obtain. The Vitoria system was able to obtain this metric by first obtaining the GPS information of the user and then querying a reverse geocoding \gls{API}, namely the OpenStreetMaps \gls{API} \cite{ReverseN98:online}. By using this method, the system was able to infer the user district, municipality, and parish. As previously stated, the GPS information was never stored due to privacy concerns, and in the case that no internet connection was available the system would discard the GPS information. Additionally, storing the name of municipalities or districts could raise privacy concerns as well, even with anonymised user identity. As such, the system anonymized the denominations obtained from the reverse geocoding system as well. Furthermore, in order to further preserve user privacy, the information anonymization was user specific. That is, even for 2 users on the same location, the stored information would be different. This allowed our system to detect if a specific user was in different locations, the duration of said stay and the frequency of those movements, but never the specific location or his/her interactions with other users

In addition, in Portugal, the municipalities were classified in 4 risk levels, ranging from normal to extreme in terms of risk of infection. In order for our system to keep track of this information as well, a back-end service was configured to maintain an updated list of Portuguese municipalities and their respective risk levels. The mobile application was then able to, prior to the anonymization of the geographical information, query this service to obtain the current risk of infection due to geographic location of the user.

Another important metric to consider due to the nature of the pandemic, is user activity levels. Due to sudden change in habits caused by confinement, most people saw their daily routines affected in terms of sleep and physical activity. As such, our system proposes a method to passively collect that data by using both Google’s Physical Activity Recognition \gls{API} \cite{ActivityRecognition:online} and the Sleep Recognition \gls{API} \cite{SleepAPI:online}. This Activity Recognition \gls{API} is able to automatically detect activities by periodically reading and processing sensor data and outputting a classification whenever the user activity changes. This \gls{API} is able to classify user activities into: running, walking, on bicycle, in vehicle, on foot, tilting, and still. On the other hand, the Sleep Recognition \gls{API} periodically returns a classification on whether or not the user is sleeping. Additionally, both \gls{API}s return a confidence on their classification, which allows the Vitoria system to post-process these classifications.

\begin{figure*}[t!]
\centering
\begin{subfigure}{.30\textwidth}
\centering
\includegraphics[width=1.8in]{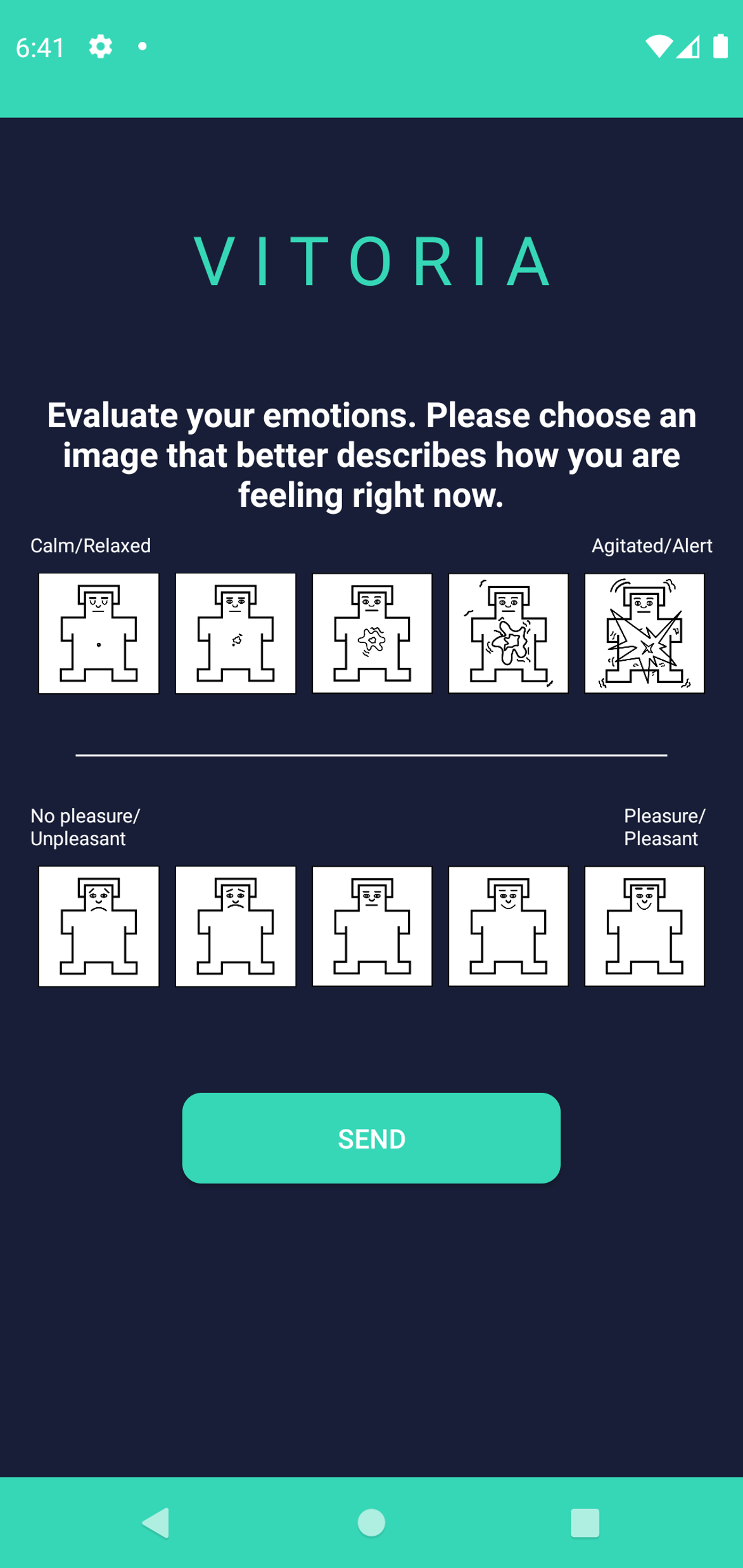}
\caption{}
\label{fig:emotional_form}
\end{subfigure}
\begin{subfigure}{.30\textwidth}
\centering
\includegraphics[width=1.8in]{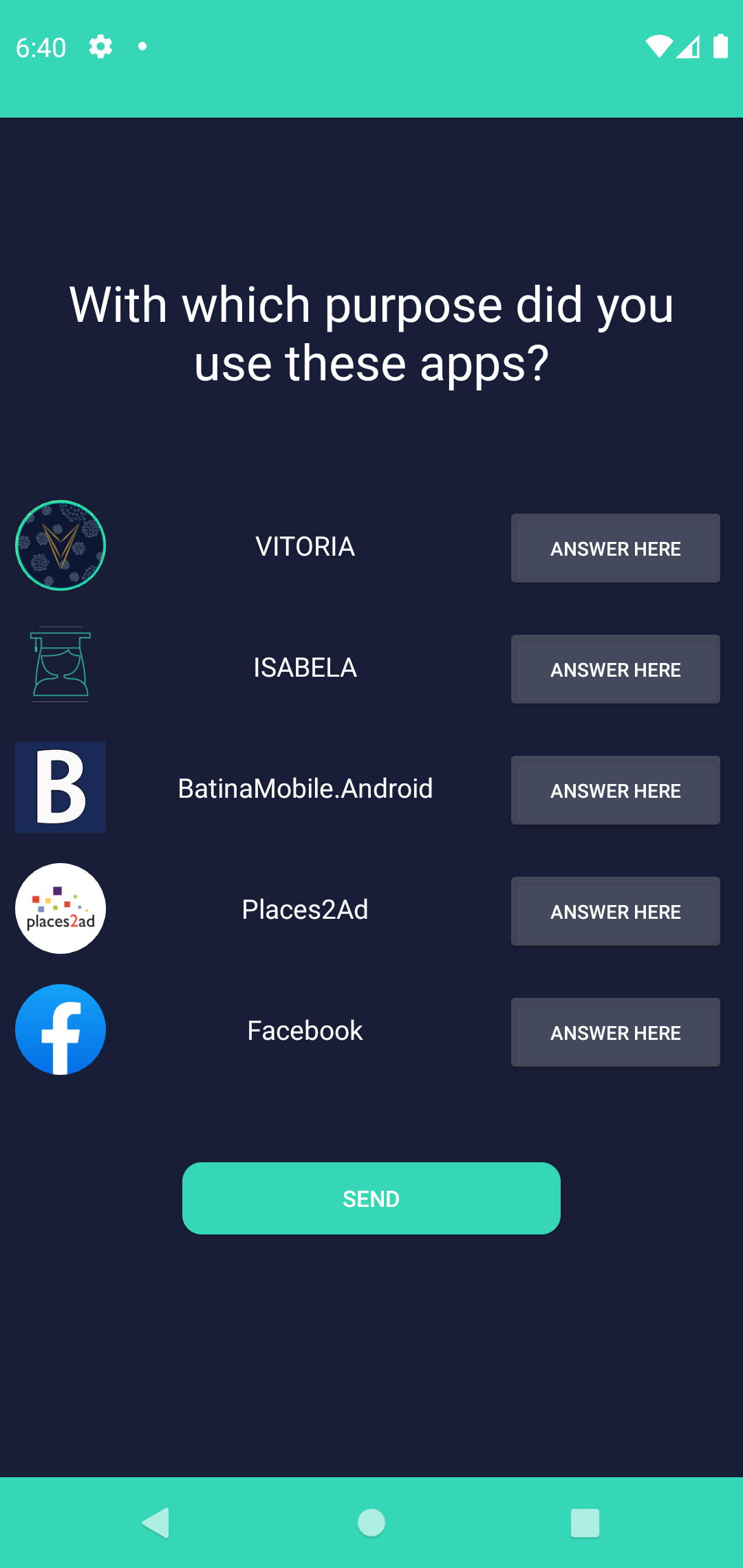}
\caption{}
\label{fig:finality_form}
\end{subfigure}
\begin{subfigure}{.30\textwidth}
\centering
\includegraphics[width=1.8in]{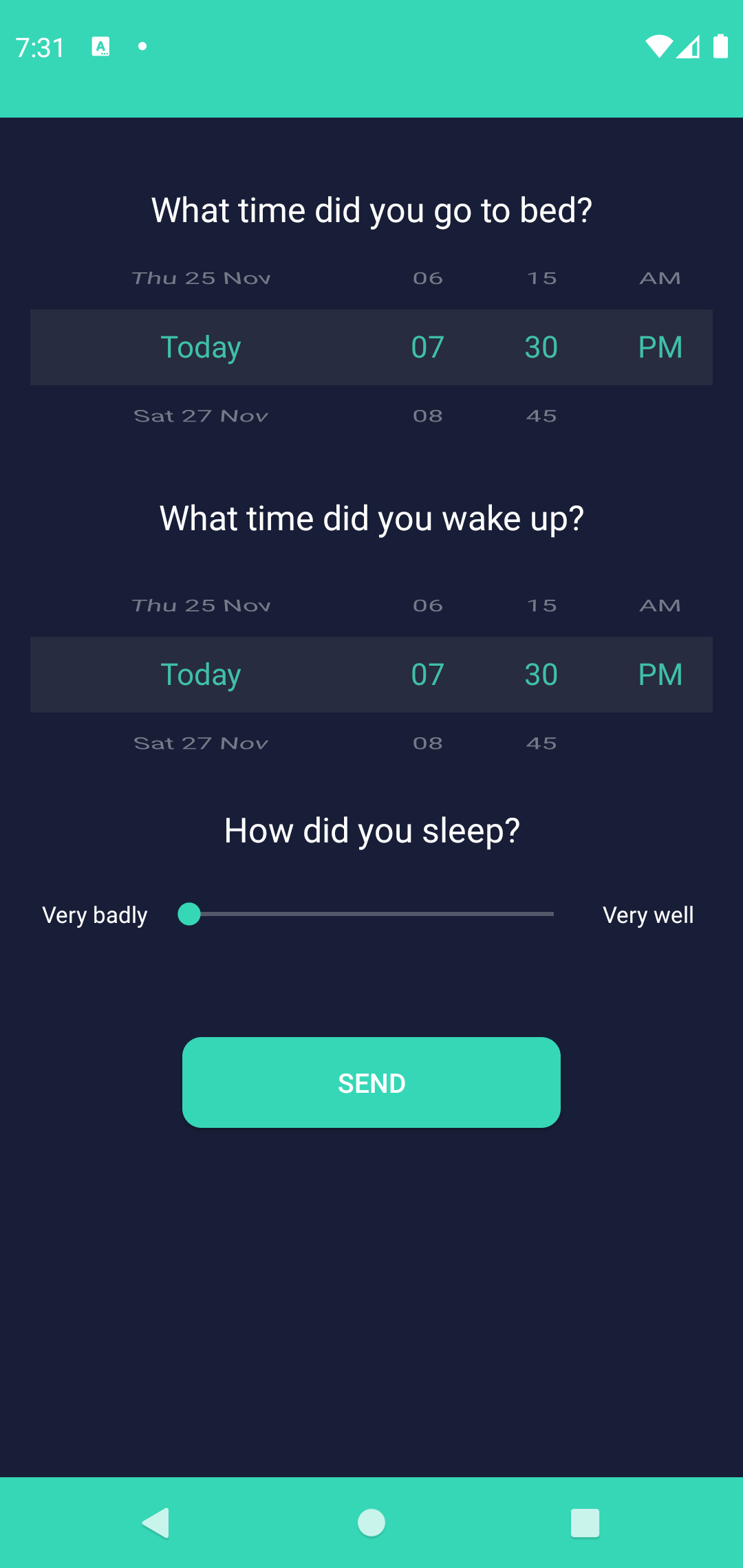}
\caption{}
\label{fig:sleep_form}
\end{subfigure}
\caption{Vitoria's Questionnaires: Emotional questionnaire using the SAM scale (a); Application finality form (b); Sleep duration and sleep quality questionnaire (c).}
\label{fig:forms}
\end{figure*}

Another metric that can be used to infer the changes on user behaviour, is the statistics of the used applications. The Android \gls{SDK} allows us to retrieve information about which applications were used, and for how long they were used. This information can, for instance, be used to infer changes in the user behaviour due to confinement, or to infer the frequency and duration of user interaction with other people through communication apps.

Smartphones can also be used to create a virtual sensor that is able to detect ambient noise. In the Vitoria system, we were interested not only in obtaining information about the user but also about their context. As such, our system uses the microphone to compute the noise amplitude in the users’ proximity. As with the other metrics, here privacy is also a major concern. As such, no audio is stored nor recorded, and the system microphone is just used to average the ambient noise in short intervals of time.

Another important aspect of people’s behaviour in the context of a pandemic is their proximity to other people and the number of people they have contact with. For this, we used the Bluetooth smartphone capabilities to detect nearby devices that could correspond to people in the user vicinity. The Bluetooth \gls{API} from the Android \gls{SDK} \cite{Bluetoot33:online} allows to differentiate the type of devices that are detected with the Bluetooth and \gls{BLE} scans. These types can be for instance, headphones, smartphones, televisions, cars, wearables, etc. Additionally, all already connected devices are filtered and, as such, we can consider that all detected headphones, smartphones, and wearables correspond to people in the proximity of the user. However, the same person could be using several of these devices. As such, we do not infer this information directly. Instead, we use this data to opportunistically ask the user to answer a questionnaire. This proximity questionnaire is explored in the next section. In addition to the type of device, the strength of the signal is also considered, to infer the user distance to the device.

The Vitoria system is also able to scan and store the WIFI information. Usually, places with a large number of people, like supermarkets, shopping centres, gyms and universities have a large number of WIFI access points. As such we believe that this information can also be used to infer when the user is in such places. In order, to maintain the user privacy, the name of the detected Access points is never stored, only the MAC address and the signal strength.

As we previously stated, in addition to the smartphone application, the Vitoria system also comprises a smartwatch application. This application serves as a mean to collect physical and physiological data from the users. The smartwatch application is capable of infer the type of activity that the user is doing, the number of steps taken and the heart rate values of the user. Although activity data is already collected from the smartphone, the smartwatch is closer to the user and can offer a more reliable insight on the user activity levels. Moreover, physiological data can also offer a valuable insight about the user’s physical and emotional states, especially the heart rate values, which have been proved to be correlated with stress levels  \cite{taelman2009influence}, for instance. Additionally, the heart rate can also be used to post-process sleep recognition data or to infer sleep quality.

In addition to the aforementioned metrics, the Vitoria system is also able to collect raw sensor data such as accelerometer data, gyroscope data, proximity data, and ambient light data. Although these metrics are not yet used, they can be used to create novel classification systems, or to improve existing ones. For instance, the values from the light sensor can be used to post-process the information from the sleep recognition \gls{API}. All of the mentioned sensors are collected from both the smartphone and smartwatch applications.

\subsection{Active Data Acquisition}
\label{sec:active_data}

Although smartphone context information and sensors data can be used to infer several aspects of the users’ daily life, when considering the users’ psychological states these kinds of metrics have yet to be validated. As such, there is a need to use more conventional methods, such as questionnaires. In the Vitoria system, we implemented several questionnaires to infer the user’s emotional state, and to complement the already mentioned data. In this section we describe the implemented questionnaires.

One of the aspects that we consider important to monitor during the pandemic context is the emotional state of people. Emotional states of humans are difficult to perceive and infer, and no approach that uses sensors has been validated as an adequate means of inferring these states. As such, we implemented a questionnaire-based approach to obtain this type of data. The Vitoria system implements the \gls{SAM} scale \cite{Bynion2020} as a questionnaire that is daily prompted to the user. The \gls{SAM} scale is an image based self-assessment scale to evaluate two dimensional emotions of the users. The questionnaire can be seen in Figure \ref{fig:emotional_form}. The two scales correspond to arousal (top) and valence (bottom). This questionnaire was prompted to the user on a daily basis and at random time between 14h-20h. The questionnaire was not released during the morning because, during this period, the user perception can still be affected by the events of the previous day.

As mentioned in the previous section, the used applications and the respective usage duration might hold important information about the user. As such, we implemented another questionnaire to additionally determine the purpose of the applications’ use. As can be seen in Figure \ref{fig:finality_form}, in this form the user is prompted to select the purpose to which he/she used each of the five most used apps during the last 24 hours. This questionnaire was prompted to the user every day at 14h. The user can select one of four major purposes: communication, leisure, research, and work. We believe these four purposes make up most of the purposes for which every application today is used. The data obtained from the user answers to this form will be explored in more detail in the section \ref{sec:app_finality}.

Another aspect that the Vitoria system aims to explore is user sleep habits. As such, the Vitoria system implements a questionnaire that prompts the users to select the time at which they went to bed, at which time did they wake up, and how they classify their sleep, as can be seen in Figure \ref{fig:sleep_form}. In addition, with the objective of being able to use this questionnaire to validate the data retrieved from the Sleep Recognition \gls{API}, the system is able to retrieve information about the sleep quality of the users. We intend to use the latter to extend the Sleep Recognition \gls{API} with the capabilities to detect the sleep quality as well.

\begin{figure*}[!t]
\centering
\begin{subfigure}{.3\textwidth}
\centering
\includegraphics[width=1.8in]{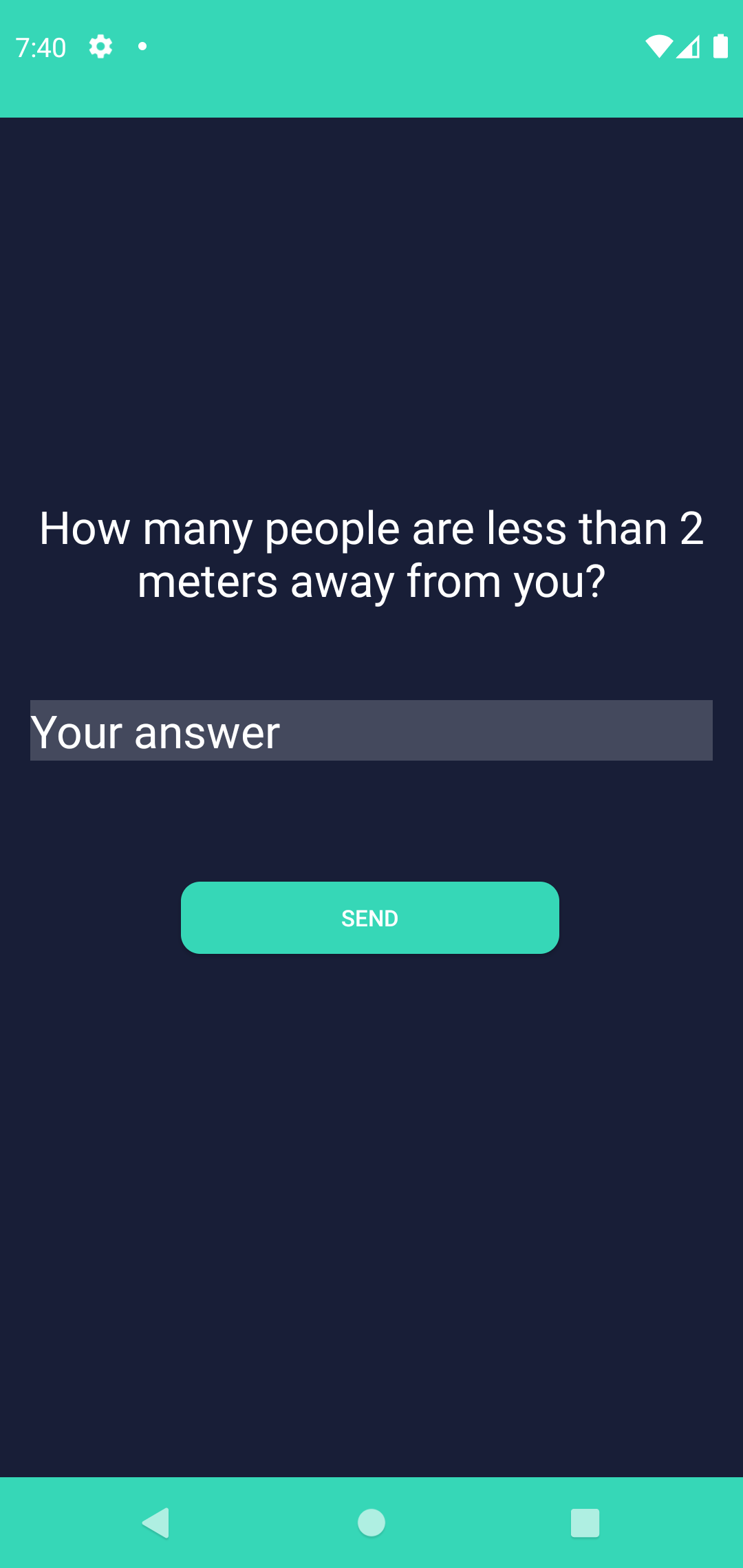}
\caption{}
\label{fig:proximity_form}
\end{subfigure}
\begin{subfigure}{.3\textwidth}
\centering
\includegraphics[width=1.8in]{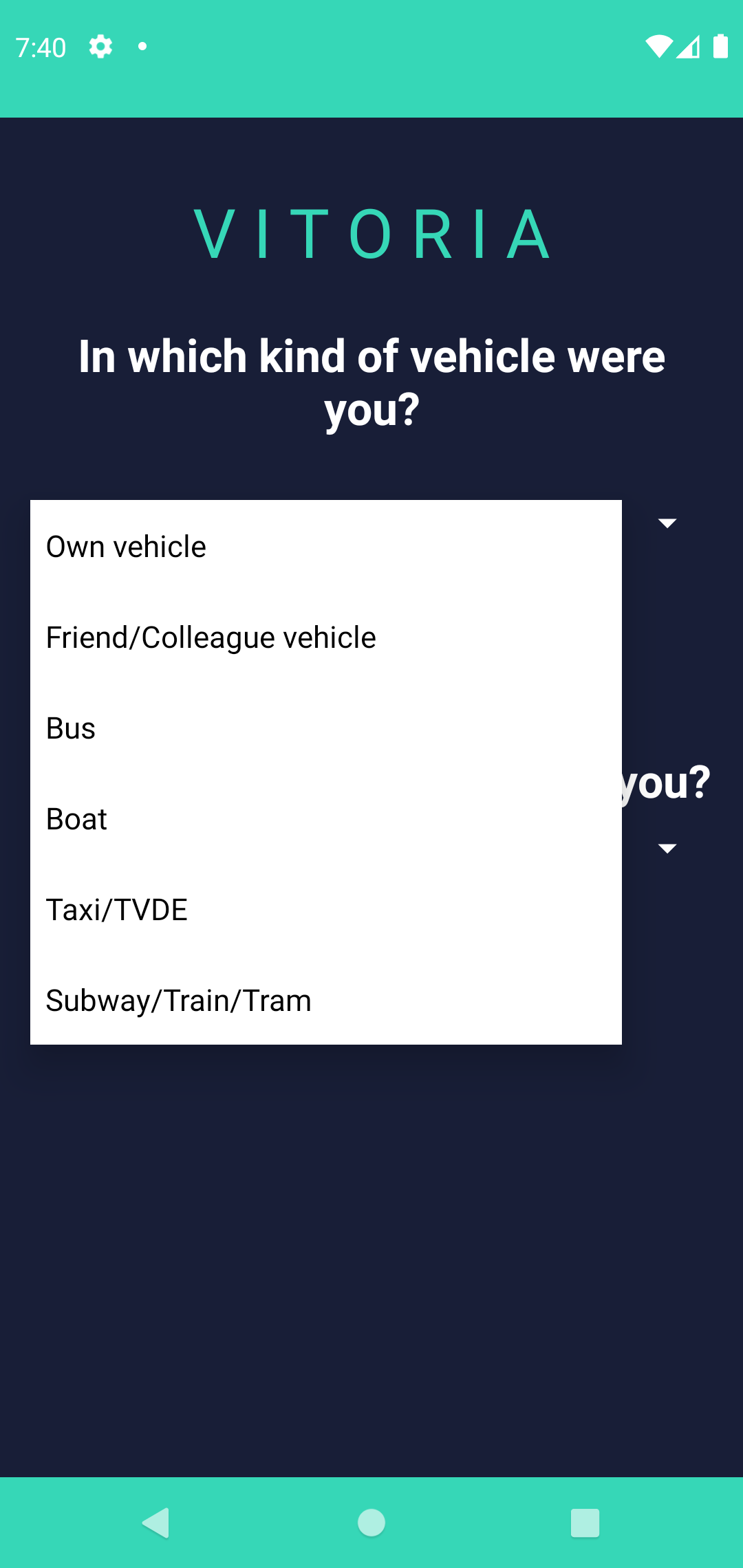}
\caption{}
\label{fig:transport_form_1}
\end{subfigure}
\begin{subfigure}{.3\textwidth}
\centering
\includegraphics[width=1.8in]{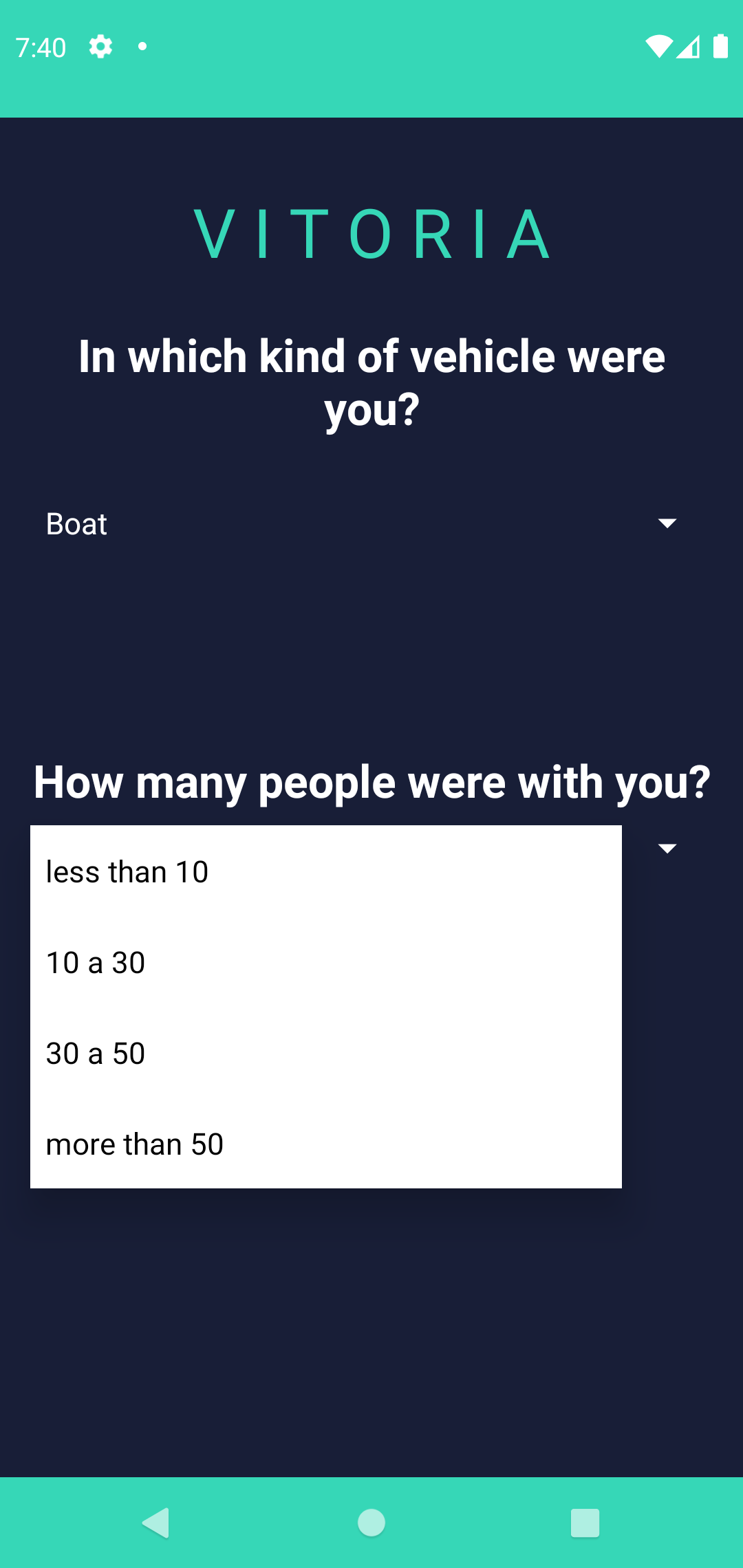}
\caption{}
\label{fig:transport_form_2}
\end{subfigure}
\caption{Vitoria's Opportunistic Questionnaires: Proximity questionnaire (a); Transport information questionnaire transport selection open (b); Transport information questionnaire with number of people selection open (c).}
\label{fig:forms_transport}
\end{figure*}

Apart from the questionnaires that are prompted to the user on a daily basis, our system is also able to show the user opportunistic questionnaires, based on their current context. One of them, is the \textit{Proximity/Contacts Questionnaire} that can be seen in Figure \ref{fig:proximity_form}. The number of contacts a person maintains is very important when considering the context of a pandemic. When the application detects more than two devices corresponding to other people, we raise the Proximity Questionnaire, The user is then prompted to answer how many persons are closer than two meters from him/her, in a numeric type answer. To prevent the application from being bothersome to use, we added a cooldown of one hour to this questionnaire.

\begin{table}[!t]
\caption{Number of persons on the surroundings options by type of transportation.}
\centering
\label{table:options_transport_form}
\renewcommand{\arraystretch}{2.15}
\begin{tabular}{|c||c||c|}
\hline
\textbf{Own car} & \multirow{2}{*}{\textbf{Bus}} & \multirow{3}{*}{\textbf{Boat}} \\
\textbf{Friend/Colleague vehicle} & \multirow{2}{*}{\textbf{Subway/Train/Tram}} & \\
\textbf{Taxi/TVDE} & &  \\                                                        \hline \hline
0 & less than 10 &less than 10 \\
1 & 10 to 20 & 10 to 30\\ 
2 & 20 to 30  & 30 to 50 \\ 
more than 2 & more than 30 & more than 50  \\
\hline
\end{tabular}
\end{table}

Additionally, another aspect that can be considered a risk factor for contamination with the Covid-19 virus is the use of public transportation or the use of shared vehicles with restricted space. One of the activities that the Activity Recognition \gls{API} is able to detect is if the user is in a vehicle. By using this feature, the Vitoria application is able to prompt the user with the \textit{Transport Questionnaire}, which can be seen in Figures \ref{fig:transport_form_1} and \ref{fig:transport_form_2}. In this questionnaire the users should indicate the transport type and the number of people in the vehicle that are close to them. As can be seen in Figure \ref{fig:transport_form_1}, the available options are: \textit{Own vehicle}; \textit{Friend/Colleague vehicle}; \textit{Taxi/TVDE}\footnote{Designation for transports from electronic platforms such as Uber or Bolt in Portugal.}; \textit{Bus}; \textit{Subway/Train/Tram}; \textit{Boat}. Depending on the type of vehicle, the options to indicate the number of people are different. The full list of options can be seen in Table \ref{table:options_transport_form}. This questionnaire is triggered every time the\textit{ ”in vehicle”} activity is detected for more than 2 minutes. If more than one transportation is used, or if there are several stops during a trip (e.g., on a bus ride) the system triggers several transportation questionnaires. However, when there are multiple transportation questionnaires, only the last one will be considered, and the remaining ones are discarded.

Although data acquired directly from the user is more reliable and is well validated in the literature, one of the main issues when collecting this type of data is the user engagement. In order to address this issue, in the Vitoria system every questionnaire is prompted to the user through the Android notification system. The user can then select each notification and will be redirected to the specific page of each questionnaire. These notifications are permanent until the user clicks or dismisses them, since the user is not required to answer them right way. Additionally, the Vitoria system triggers a notification every day at 16 hours to remind the user to answer any unanswered questionnaires. When the user clicks that notification, he/she is redirected to a page where all unanswered questionnaires are listed. The user can then select each questionnaire and answer them. The pending questionnaires have a validity of 24h after which they are deleted and are no longer shown in this page.

\subsection{Feedback System}
\label{sec:feedback}

Another aspect that the Vitoria system aimed to explore was how the use of active feedback would affect user behaviour. In order to support that, the system implements on screen personalized information feedback. As discussed in section \ref{sec:backend_achitecture}, the \textit{Feedback Grante}r module was created to deal with feedback generation and distribution in the Vitoria system. This module is able to divide the users into two feedback groups, namely the control group and the active feedback group (with a 50-50\% ratio).

\begin{figure}[t!]
\centering
\includegraphics[width=1.8in]{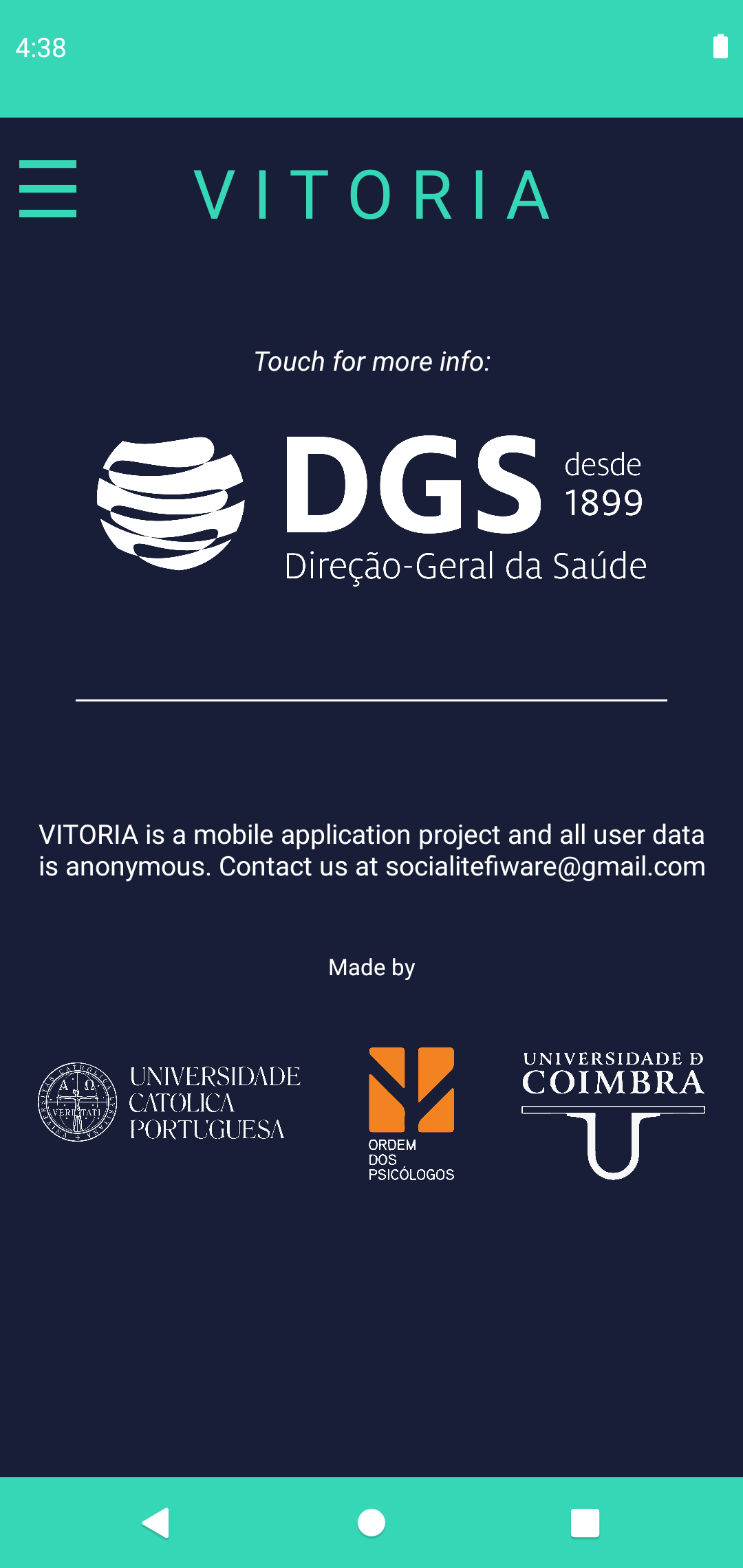}
\caption{Vitoria Main screen without feedback.}
\label{fig:no_feedback}
\end{figure}

\begin{figure*}[!t]
\centering
\begin{subfigure}{.3\textwidth}
\centering
\includegraphics[width=1.8in]{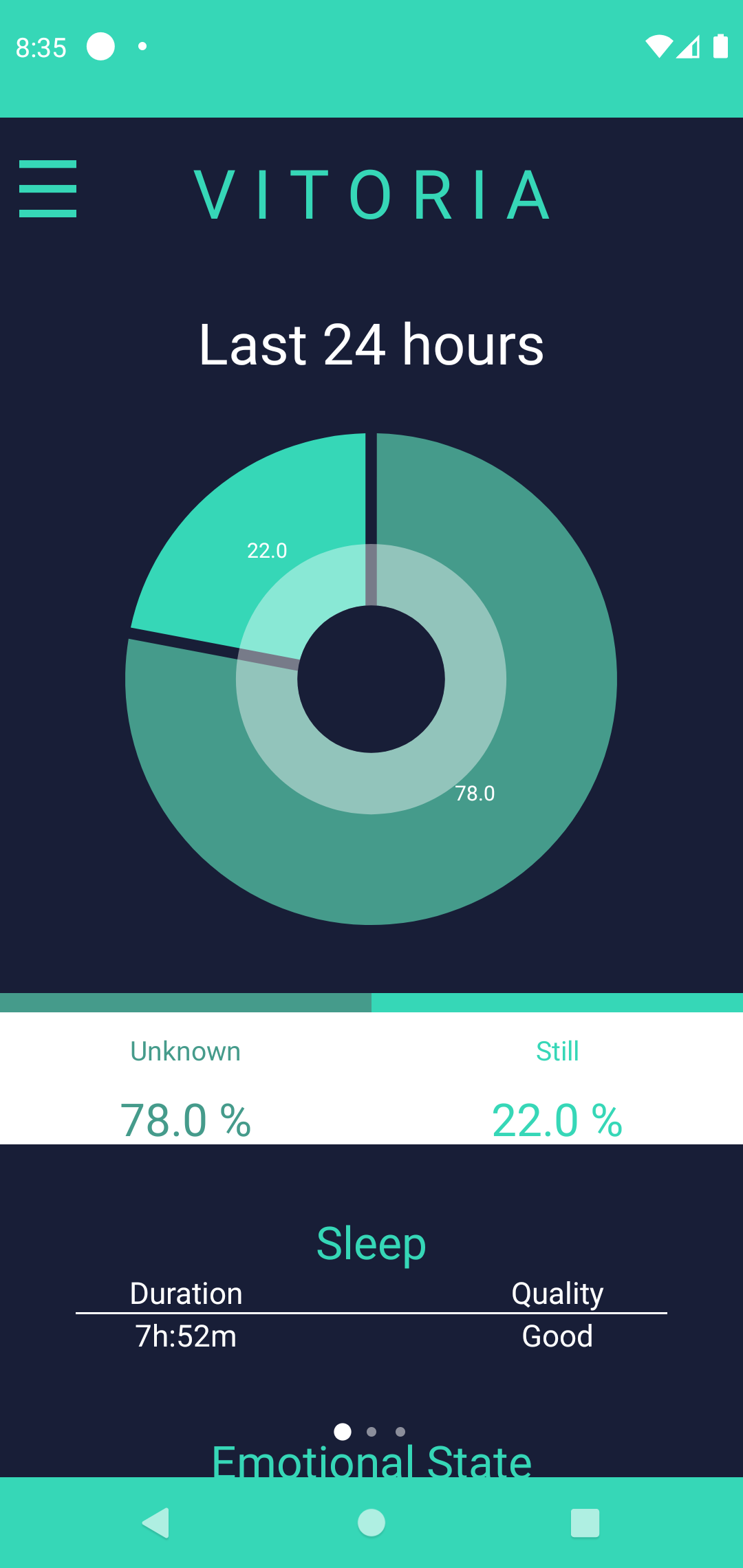}
\caption{}
\label{fig:feedback_control_1}
\end{subfigure}
\begin{subfigure}{.3\textwidth}
\centering
\includegraphics[width=1.8in]{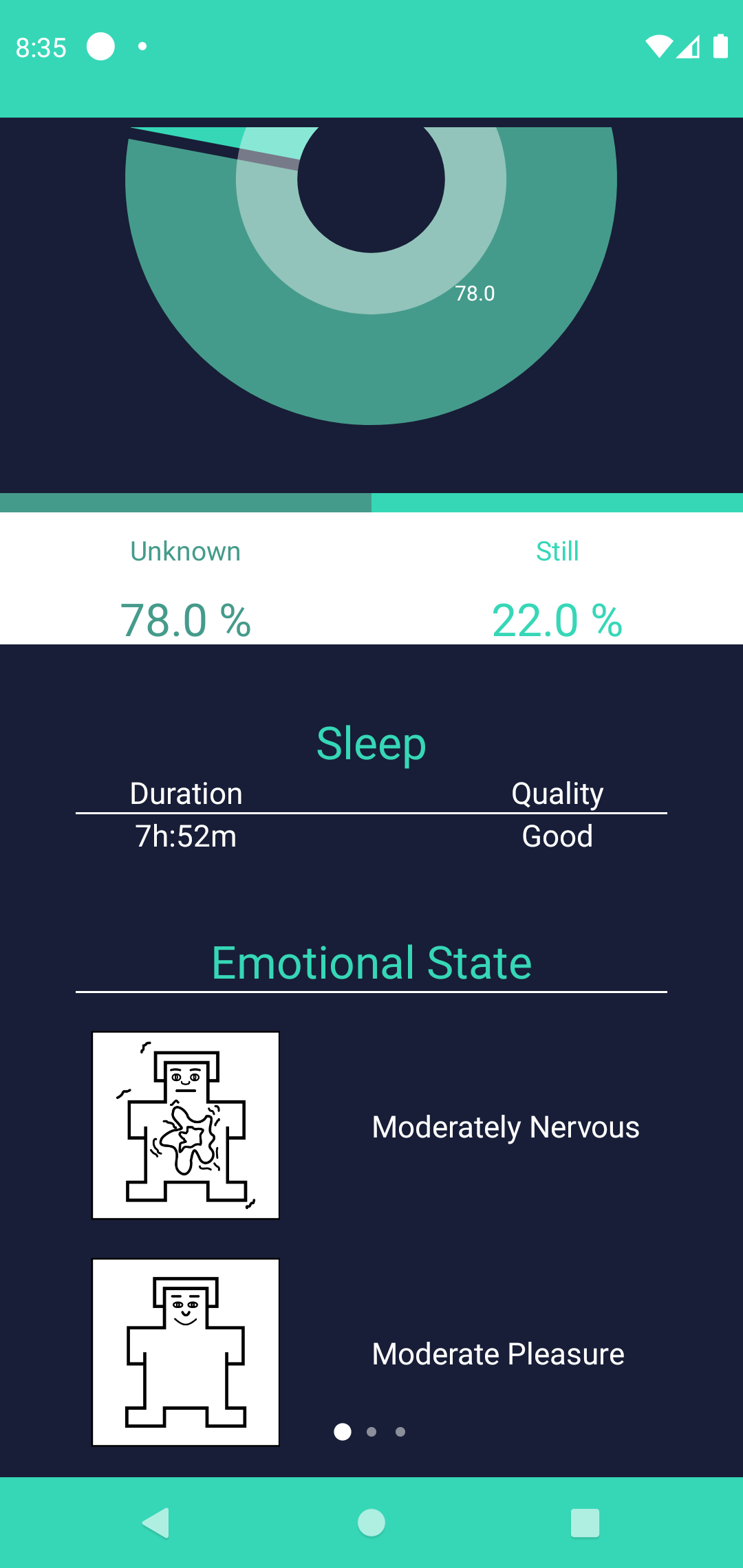}
\caption{}
\label{fig:feedback_control_2}
\end{subfigure}
\begin{subfigure}{.3\textwidth}
\centering
\includegraphics[width=1.8in]{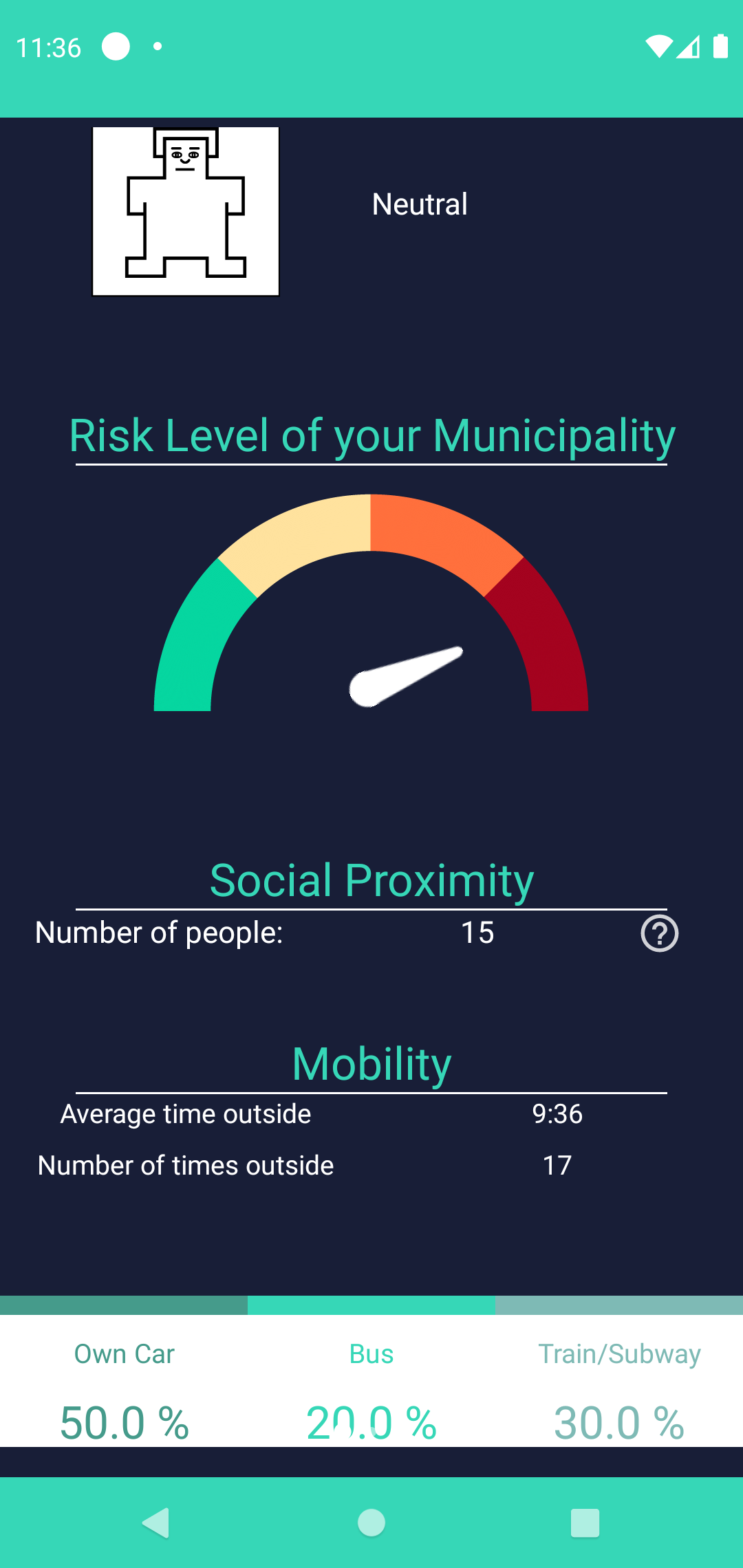}
\caption{}
\label{fig:feedback_active}
\end{subfigure}
\caption{Vitoria's Feedback layouts: Control's group Feedback  (a) \& (b); Specific Active Group Feedback (c).}
\label{fig:feedback}
\end{figure*}

Additionally, the field study presented in this paper was divided into 2 periods: a period without feedback (default measurement), and a period with feedback. In the first period, none of the two user groups receive any feedback. The application layout for this period can be seen in Figure \ref{fig:no_feedback}. In the second period both user groups received feedback. However, the feedback differed based on the user group, as can be seen in Figure \ref{fig:feedback}. The differences between the groups are presented in Table \ref{table:feedback_options}. Additionally, users were always presented with the feedback for 3 intervals of time, namely: the last 24 hours, the last 4 days, and the last 8 days. These intervals of time were the same for both feedback groups. Additionally, the feedback information was recalculated every 30 minutes.

\begin{table}[!t]
\caption{Feedback differences between groups.}
\centering
\label{table:feedback_options}
\renewcommand{\arraystretch}{2.15}
\begin{tabular}{|l|c|c|}
\hline
& \multirow{2}{*}{Control Group} & \multirow{2}{*}{Active Feedback Group} \\ 
& & \\ \hline
Physical Activity & X & X  \\ \hline
Sleep  & X & X \\ \hline
Emotional State& X & X \\\hline
Municipality Risk Level && X\\ \hline
 Social Proximity & & X\\ \hline 
 Mobility quantity & & X \\ \hline
 Mobility duration & & X \\ \hline
Mobility type &&X\\ \hline

\multirow{3}{*}{Intervals} & \multicolumn{2}{c|}{last 24 h}                   \\
                           & \multicolumn{2}{c|}{last 4 days}              \\ 
                           & \multicolumn{2}{c|}{last 8 days}    \\ \hline
\end{tabular}
\end{table}

\begin{figure*}[!ht]
\centering
\begin{subfigure}{.45\textwidth}
\centering
\includegraphics[width=1.8in]{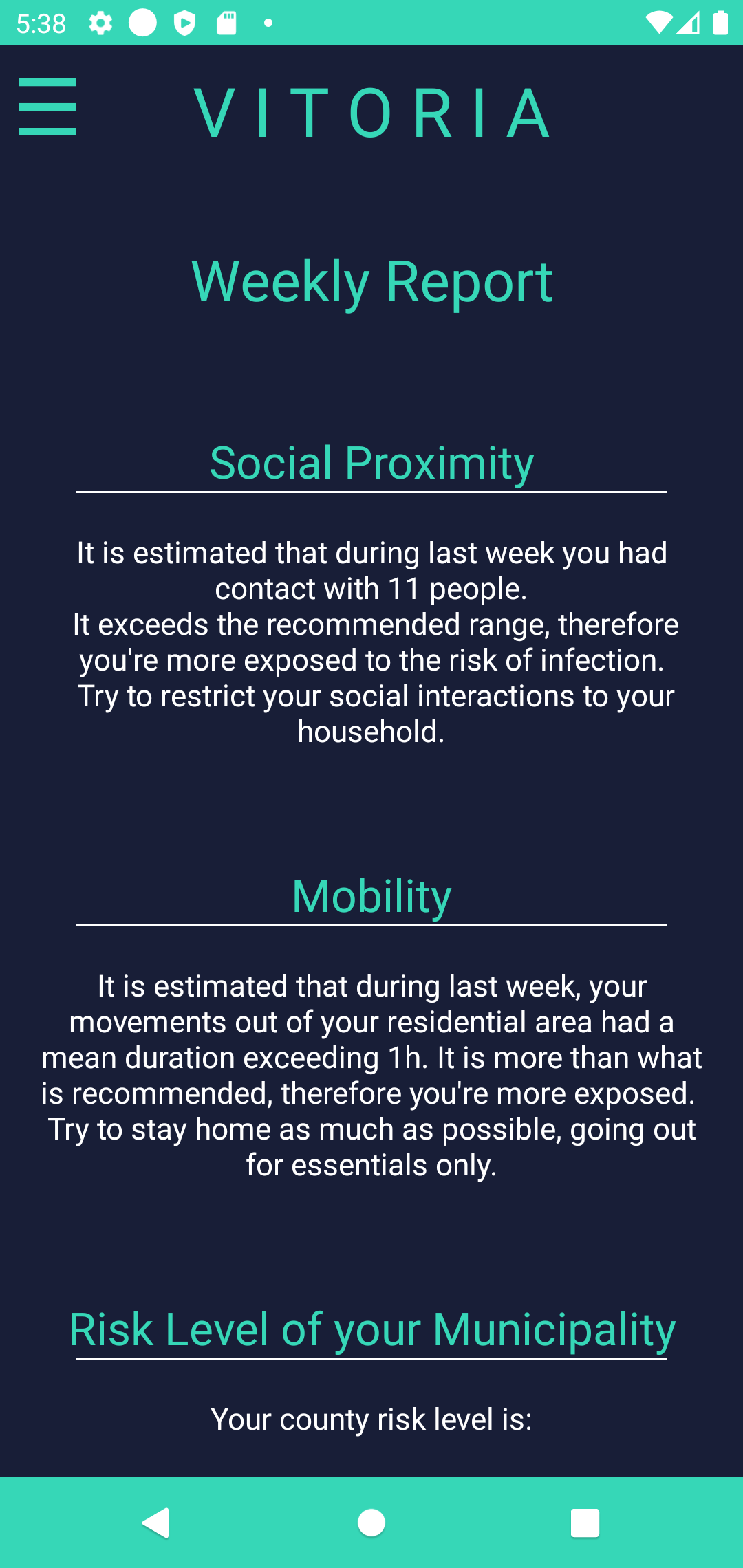}
\caption{}
\label{fig:feedback_weekly1}
\end{subfigure}
\begin{subfigure}{.45\textwidth}
\centering
\includegraphics[width=1.8in]{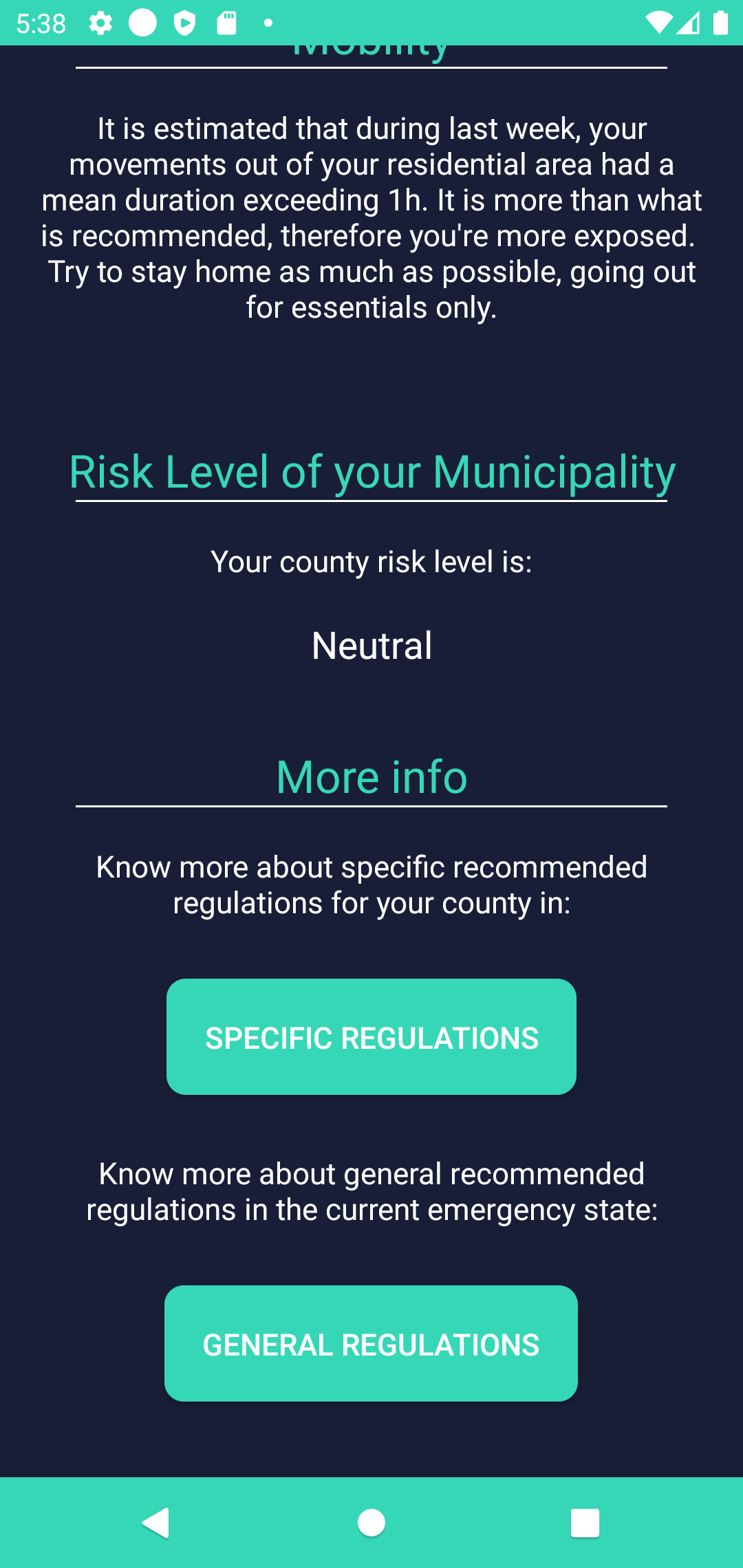}
\caption{}
\label{fig:feedback_weekly2}
\end{subfigure}
\caption{Vitoria's Full Weekly Feedback}
\label{fig:feedback_weekly}
\end{figure*}

As we can see in the table, both groups received the three first types of feedback, namely, sleep information, emotional state information, and physical activity information. The physical activity feedback is generated based on the data provided by the Activity Recognition \gls{API}, and this information specifies the percentage of time spent by the user in each activity, as can be seen in Figure \ref{fig:feedback_control_1}. Sleep feedback is generated from the collected sleep recognition data, and from the sleep report filled daily by the user. This information is shown to the user as the mean time spent sleeping and the mean quality of sleep (i.e., very bad, bad, neutral, good and very good), as can be seen in Figure \ref{fig:feedback_control_1}. The Emotional state information is generated from the daily emotional questionnaire. This feedback is presented using the same figures from the \gls{SAM} questionnaire, as can be seen in Figure \ref{fig:feedback_control_2}.

As we can see in Table \ref{table:feedback_options}, the active feedback group received more information than the control group. This information was directly related to the SARS-Cov-2 risk of infection. For instance, the users on the active feedback group received information about their municipality risk level, based on local infections and incidence rates. As explained before, in section \ref{sec:data_aquisition}, this was done by comparing the risk level obtained from the reverse geolocation with their discrete location. As can be seen in Figure \ref{fig:feedback_active}, this information is presented to the user in the form of a gauge divided into 4 sectors that represent the 4 denominations used by the Portuguese authorities (i.e., moderated, high, very high, and extremely high).

Additionally, the users from the active feedback group also received information about the mean number of possible contacts that they had in each specific time interval. This information was generated based on the answers to the \textit{proximity questionnaire}, and was presented to the users in the form of a numeric value. Users also received information about their mobility, namely the mean time the users spent outside their home, and the total number of times that they went outside. In Figure \ref{fig:feedback_active} we can see an example of this feedback for an 8-day time interval.

As we stated before, some means of transportation can lead to a higher risk of infection due to higher proximity to other people. As such, in addition to the previously mentioned feedback, the application also shows the percentage of time spent on specific means of transportation, as can be seen in Figure \ref{fig:feedback_active}. This percentage is calculated based on the \textit{transportation questionnaire} that the application prompts the user to fill whenever the \textit{"in vehicle"} activity is detected for more than 2 minutes, by the Activity Recognition \gls{API}.

Furthermore, in addition to the feedback explained so far, the users also receive weekly feedback with the aggregated data for that week. Although the users already receive feedback for the last 8 days, that information is recalculated every 30 minutes, and, as such, is continuously changing. On other hand, in the weekly report, the information is static until the users receive the next weekly report. The report was set to be sent every Saturday at 21 hours and this feedback can be seen in Table \ref{table:weekly_feedback}.

Starting with the \textit{Control Group} we can see that the feedback given to this group was generalist, with only two of the four items in the weekly feedback items. The last item in Table \ref{table:weekly_feedback} was common for both groups, with information about the current measures implemented by the Portuguese authorities. The penultimate item of the table was also given to both groups. However, only the group with active feedback received information about their respective municipality risk, while the control group only received the general guideline.

The Active Feedback group received two more items of feedback, namely the feedback about the number of contacts and the mobility outside of the home area. As we can see in Table \ref{table:weekly_feedback}, these two items are also divided into positive reinforcement and negative reinforcement. That is, whenever the user meets the recommendation from the health authorities, they receive a positive message; whenever they fail to meet the recommendations, they receive a negative one. In the case of the number of contacts, the recommendation was that users would limit the number of contacts to 10 or less people. And in the case of mobility outside of their home area, the recommendation was that it should be limited to a minimum, which led us to choose a threshold of one hour. 

As explained before, the users received a notification, every Saturday at 21 hours, which redirected them to the weekly feedback page in the mobile application. This information could then be accessed over the course of the following week, until a new notification was received, and the information was updated. This information was presented as shown in Figure \ref{fig:feedback_weekly}, with the control group only having access to the ”More Info” part of the layout.

\begin{landscape}

\begin{table}[t!]
\centering
\caption{ Weekly feedback for the Actuation groups}
\label{table:weekly_feedback}
\renewcommand{\arraystretch}{1.25}
\begin{tabular}{|c|c|c|}
\hline
\multirow{5}{*}{\textbf{Control Group}}& \multicolumn{2}{c|}{\multirow{3}{*}{\textbf{Active Feedback Group}}}\\
& \multicolumn{2}{c|}{}\\
& \multicolumn{2}{c|}{}\\
\cline{2-3} 
& \multirow{2}{*}{\textbf{Positive Reinforcement}} & \multirow{2}{*}{\textbf{Negative Reinforcement}}\\
& &\\
\hline
- - - - -  & \begin{tabular}[c]{@{}c@{}}\textbf{\textit{If social proximity below 10 people in the last week:}}\\ "We estimate that in the last week you had contact with \\\textbf{\textless{}Number of Contacts\textgreater} people, what is within \\ what is recommended.\\ Congratulations! Keep doing what you're doing,\\ limiting social contacts to your household."\end{tabular} & \begin{tabular}[c]{@{}c@{}}\textbf{\textit{If social proximity exceeds 10 people in the last week:}}\\ "We estimate that in the last week you had contact with \\ \textbf{\textless{}Number of Contacts\textgreater} people.\\ Because it is a greater number than what is recommended,\\  you are more exposed to the risk of infection by the coronavirus.\\ Limit your contacts to members of your household as much as possible."\end{tabular} \\
\hline
- - - - - & \begin{tabular}[c]{@{}c@{}}\textbf{If the user had low mobility} \\ \textbf{\textit{(less than 1 hour outside your home area on average)}}:\\"We estimate that in the last week your trips outside \\ your area of residence had an average duration of less than 1h,\\  which is within what is recommended.\\ Congratulations! Please continue to stay at home except for essential travel."\end{tabular} & \begin{tabular}[c]{@{}c@{}}\textbf{If the user had high mobility} \\ \textbf{\textit{(1h or more outside the home area on average)}}:\\"We estimate that in the last week your trips outside\\  your area of residence had an average duration of more than 1 hour.\\ Because it is a greater number than what is recommended, \\ you are more exposed to the risk of infection by the coronavirus.\\ Try to stay at home as much as possible, except for essential trips."\end{tabular} \\

\hline
\begin{tabular}[c]{@{}c@{}}"Learn more about the specific measures\\ recommended for the municipality\\ where you live at:\\\textbf{\textit{https://covid19estamoson.gov.pt/}}"\end{tabular} & \multicolumn{2}{c|}{\begin{tabular}[c]{@{}c@{}}"The risk level in the municipality where you live is \\ \textbf{\textit{\textless{}Municipality Risk Level\textgreater{}}}. Learn more about the specific measures recommended\\ for the municipality where you live,\\  at: \textbf{\textit{https://covid19estamoson.gov.pt}}"\end{tabular}}  \\
\hline

\multicolumn{3}{|c|}{\multirow{3}{*}{\begin{tabular}[c]{@{}c@{}}"Learn more about the general measures recommended to follow in the current state of emergency: \\ \textbf{\textit{\textless{}Link for current measures\textgreater}}"
\end{tabular}}} \\
\multicolumn{3}{|c|}{}\\
\multicolumn{3}{|c|}{}\\
\hline
\end{tabular}
\end{table}
\end{landscape}

\begin{figure*}[ht]
\centering 
\includegraphics[width=7in]{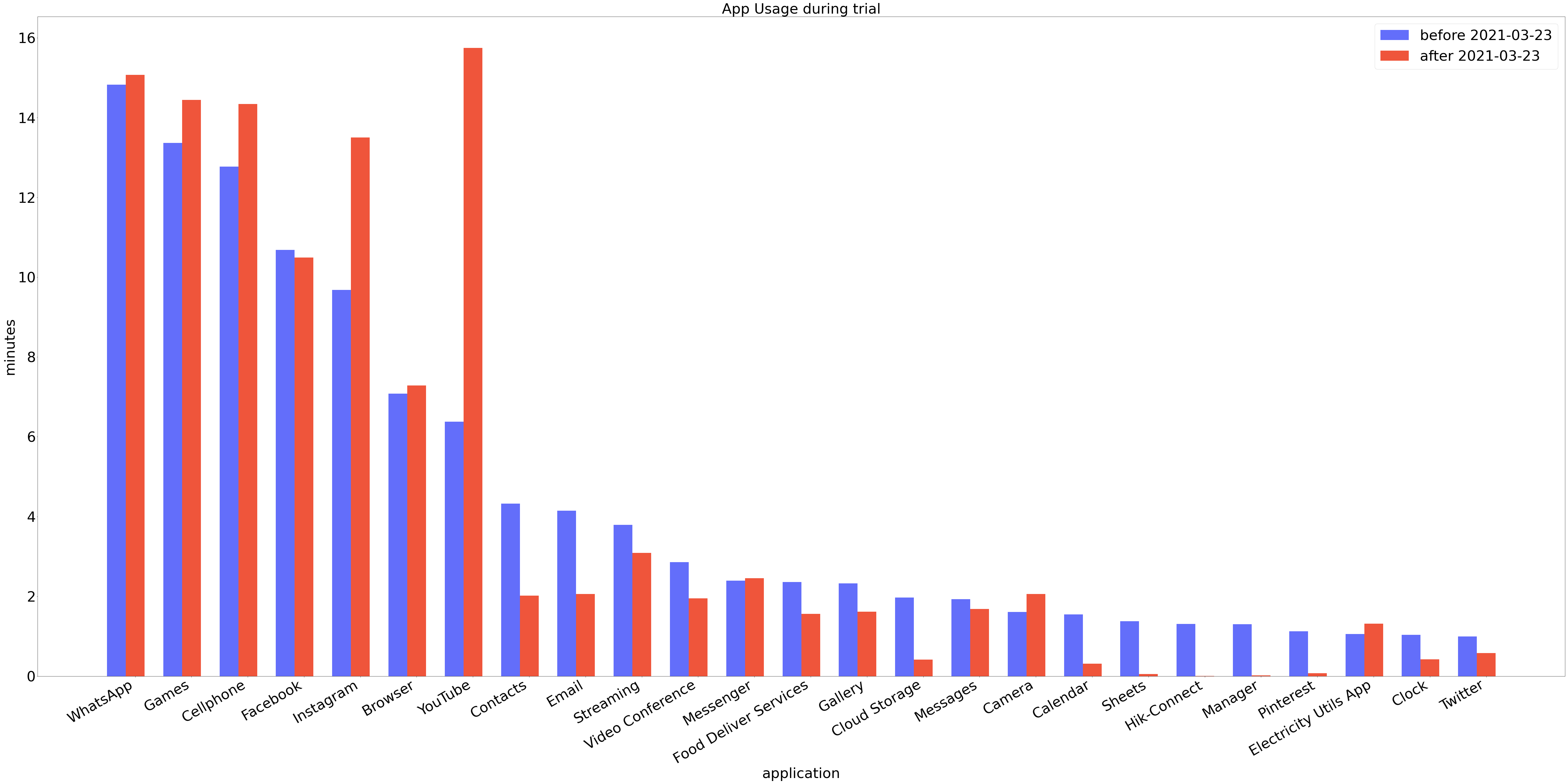}
\caption{Application usage during the 2 halves of the trial.}
\label{fig:app_usage_both_periods}
\end{figure*}

\newpage

\begin{table*}[!t]
\caption{Covid-19 related events in Portugal from the 8th of January to the 23rd of May 2021}
\label{Table:events_covid}
\centering
\renewcommand{\arraystretch}{2.2}
\begin{tabular}{|l||l|c|}
\hline
\textbf{Date} & \textbf{Event} & \textbf{Perception}\\
\hline
08-01-2021 & The number of daily cases of infections by COVID-19 surpasses the 10000 & \textbf{-}\\
\hline
12-01-2021& The Pandemic numbers increase in all age groups & \textbf{-}\\
\hline
14-01-2021& The Government announces new measures to control the proliferation of the pandemic & \textbf{-} \\
\hline
18-01-2021& The new measures come into effect & \textbf{-}\\
\hline
21-01-2021& New containment measures and the government closes educational institutions & \textbf{-} \\
\hline
\multirow{2}{*}{28-01-2021}& The barrier of 300 new deaths from COVID-19 was surpassed and the record for the highest number of new cases & \multirow{2}{*}{\textbf{-}}\\
&  ever is registered with 16432 new cases &\\
\hline
09-02-2021& Epidemic shows decreasing trend & \textbf{+}\\
\hline
12-02-2021& Government announces that \textit{"the current level of confinement will be maintained during the month of March"} & \textbf{-}\\
\hline
22-02-2021& The Minister of Health highlights the decreasing trend, but remembers that \textit{"nothing is certain"} & \textbf{+} \\
\hline
01-03-2021& 35\% of people over 80 years old have already been vaccinated & \textbf{+}\\
\hline
03-03-2021& A year of pandemic in Portugal is marked & \textbf{-}\\
\hline
08-03-2021& Study reveals perceptions and concerns of the Portuguese over a year of pandemic & \textbf{-}\\
\hline
12-03-2021& Government unveils gradual reopening plan until May 3rd & \textbf{+}\\
\hline
13-03-2021& Study sets red lines for intervention in COVID-19 epidemic & \textbf{-}\\
\hline
15-03-2021& Portugal suspends administration of Astrazeneca vaccine; The 1st phase of deconfinement & \textbf{+} \\
\hline
22-03-2021& Portugal resumes administration of Astrazeneca vaccine & \textbf{+}\\
\hline
26-03-2021& More than 1 million Portuguese were vaccinated with the first dose of the Covid-19 vaccine & \textbf{+}\\
\hline
\multirow{2}{*}{05-04-2021}& The 2nd phase of deconfinement begins (e.g., face-to-face teaching in the 2nd and 3rd cycles; & \multirow{2}{*}{\textbf{+}}\\
&reopening of restaurants' outside spaces) &\\
\hline
13-04-2021& 
Portugal registers an increase in incidence between 0 and 9 years old& \textbf{-}  \\
\hline
19-04-2021& The 3rd phase of deconfinement begins (e.g., face-to-face teaching in high-school and higher education; reopening of restaurants) & \textbf{+}\\
\hline
23-04-2021& Opening the schedule of vaccination for users over 65 years of age & \textbf{+}\\
\hline
26-04-2021& Registered a day without deaths by COVID-19 in Portugal & \textbf{+}\\
\hline
03-05-2021& The 4th phase of deconfinement begins (e.g., no commercial time restrictions; outdoor events w/ low capacity) & \textbf{+} \\
\hline
\multirow{2}{*}{11-05-2021} & Sporting becomes national football champion, in a match played at Sporting's stadium in Lisbon, & \multirow{2}{*}{\textbf{-}}\\
& people take to the streets to celebrate &\\
\hline
\multirow{2}{*}{23-05-2021} & SC Braga becomes wins the national cup of football, in a match played at the Jamor stadium in Lisbon, &\multirow{2}{*}{\textbf{-}} \\
&people take to the streets to celebrate & \\                  
\hline
\end{tabular}
\end{table*}

\newpage

\section{Trial and Results}
\label{sec:trials}

As previously stated, a study was developed and conducted in partnership with the Portuguese National Health System, that included the use of online questionnaires and the use of the Vitoria application for smartphone and smartwatch. To evaluate the use of this application, a three and a half months trial was performed from the 1$^{st}$ of February to the 133$^{th}$ of May, in Portugal. Participants were from several locations in Portugal, although predominantly from the district area of Lisbon. The study included 19 participants from all age groups and genders. From those 19 participants, we were only able to obtain data from 14 subjects, as some of them forgot to start the application after booting the smartphone or voluntarily stopped the application. Furthermore, some users did not reply to all the daily questionnaires, presumably because they forgot, which led to less samples of data.

During the trial, several Covid-19 related events occurred in Portugal, of which the most relevant ones are listed in Table \ref{Table:events_covid}. As we can see in the table, there were four deconfinement phases that gradually removed the restrictions. The trial explored the data in two periods, in co-occurrence with the changes at the national level.

Due to the relatively low number of participants in the trial, we did not evaluate the data of each individual participant. Instead, we only explored the data in terms of the mean value reported by day and by user. Additionally, although one of our objectives was to evaluate the implemented feedback, lack of data also invalidated such analysis, as dividing the dataset into two parts would further reduce the size of the available data.

The low number of users prevents general conclusions about a bigger population. As such, in this paper we aim to present preliminary results of our system, as a proof of concept, that should be further tested with larger samples. Namely we will explore the usage of different applications and the possible correlation of some of the acquired data with the official numbers reported for the pandemic in Portugal.

\subsection{App usage during the pandemic}

The types of used applications and the respective usage duration can be used to derive important information about user behaviour. For instance, the preference for applications that are mostly used for leisure as opposed to those that are related to work, could indicate that a user mostly uses his/her phone to relax. In this section we analyse the data retrieved directly from the Android \gls{SDK} application usage statistics, and the data retrieved from the user responses to the \textit{"application purpose form"}.

Android \gls{SDK} allows us to retrieve statistics concerning the usage of applications. These statistics come in the form of the total time that an application was used, in milliseconds. In the case of the Vitoria system, we were only interested in the time a user actively used an application. As such, we only retrieved the time an application was in the foreground, that is the amount of time an application was actively on the screen.

In the data collected from the participants it was possible to identify more than 700 different applications. Some of these were applications from the smartphone manufacturers, and others represented Smartphone capabilities like storage of Contacts or Messages. Due to the heterogeneity of the systems, and to the large number of applications on offer nowadays, it was not expected that all users had the same installed applications, and this was in fact exactly what happened. To deal with this issue, in the case of participants that did not have a certain application installed, a zero-minute mean value of usage for that application was considered. Additionally, similar applications were aggregated by categories.

The first aspect that we tried to determine was if the evolution of the pandemic, in Portugal, co-occurred with changes in the use of smartphone applications. We can see in Table \ref{Table:events_covid} that the first phase of deconfinement in Portugal started on the 15$^{th}$ of March. As such, we choose to divide the data into two groups, before the 23$^{rd}$ of March 2021 and after that, since that was the day that marked the middle of the study and was very close to the 1$^{st}$ deconfinement date.

In Figure \ref{fig:app_usage_both_periods} we can see the 25 most used applications in both periods, i.e., before the 23$^{rd}$ of march and after that date. The usage is expressed as the mean value of time spent using an application by day and by user, in minutes. It is also possible to see in the figure that the most used applications are messaging apps (WhatsApp, Messenger), gaming apps, cell phone apps, social network apps (Facebook, Instagram, Twitter), and streaming apps like Netflix, HBO, Stremio, etc. In the set of most used applications, it is also possible to see some applications that are mostly related to work, such as email applications, cloud storage or video conferencing applications.

Some of the applications were aggregated into classes of applications. For instance, different games were all aggregated into the Games category. The same happened for applications for Streaming, Email, Contacts, Browser, Cell phone, Cloud Storage, Camera, and Video Conferencing. We choose to do so for three reasons. Firstly, most smartphone manufacturers include different applications for some of those functionalities, such as for cell phone applications and Camera. Secondly, it was not our goal to evaluate the most used application but rather evaluate the type of usage. For instance, it does not matter if a person uses the Netflix application or the Amazon Prime application, since both of them serve the same purpose (i.e., content streaming). The third reason is to eliminate biases and heterogeneity between users. For instance, in the games class we found more than 50 different games, and it would be unreasonable to individually compare their usage, since most users do not have the same games installed.

Concerning the mean total time spent on apps, the times in both periods are very similar, with 178 minutes in the periods before the 23$^{rd}$ of March and 151 minutes after that date, which is in line with the mean usage times in other studies \cite{andone2016age}. Furthermore, the top 25 applications presented in Figure \ref{fig:app_usage_both_periods} account for approximately 2/3 of the total usage. We can see that although the 7 most used apps seem to all increase in usage in the second period, the mean time spent using applications on the Smartphone is smaller for this period, which indicates that the increase was indeed in these specific applications and not in the use of the smartphone.

The Instagram app, like most social networking applications, can be used to either consume content or publish new content. There was an increase in the use of the Instagram application in the second half of the study. This was accompanied by an increase in the use of the camera application as well. Furthermore, the increase is of similar order, with an increase of 40\% in the use of the Instagram app and an increase of 68\% for camera applications, which could indicate that, in fact, the use of the Instagram application was more related to publishing content. This could also be related to people leaving their homes more often and finding more ”interesting things” to capture in photos. Other studies have explored the relation between different types of use of this application and depressed states \cite{frison2017browsing}. As such, in future studies it would be interesting to monitor the type of use of this and similar applications as they could elicit more information about the users’ mental states.

YouTube was the application with the highest increase of all apps, amounting to more than the double. However, due to the low number of participants in the study, it is likely that this was an outlier, i.e., that one person, or a small subset of people, were responsible for this increase. We then verified that the YouTube application was mostly used by two users, and in fact these users had a big increase from an average 30 minutes to 1 hour and 30 minutes in the second period. Although, as we suspected, the increase was indeed affected by the small number of participants, the trend in the increase of YouTube use is still valid, since these users were also the users which presented the highest usage time in the first period.

There was also a decrease in applications related with work tasks, such as email, video conferencing apps (Teams, Skype, Zoom), cloud storage apps, the calendar, and even Sheets. This could point to one of two situations: 1) the use of remote work apps decreased due to the unfavourable work situation (i.e., higher unemployment); 2) people took some time to get used to the remote work situation and, with time, replaced the use of mobile applications with the use of more ”work friendly” ways to use such functionalities, such as using their laptop. It is also important to refer that, with the technological evolution, it is more frequent for people to use social networks as a work tool. Additionally, in Portugal it is very common for some companies to use WhatsApp as a tool for communication within the company. As such, it could be hard to infer if the purpose of the usage of certain application was related to work or not. In the next subsection, we explore the answers of the users to the application purpose questionnaire, which can be used to better infer these aspects.

Additionally, due to privacy reasons, no demographic information from the users was collected, and so we do not have access to information such as age or gender of each participant in the study. It has already been proven in past studies that the use of different applications is highly correlated with users' age and even gender \cite{andone2016age}. Making future studies with more users, and access to demographic information, may also provide new information related to the use of different types of applications.

\begin{figure*}[!ht]
\centering 
\includegraphics[width=7in]{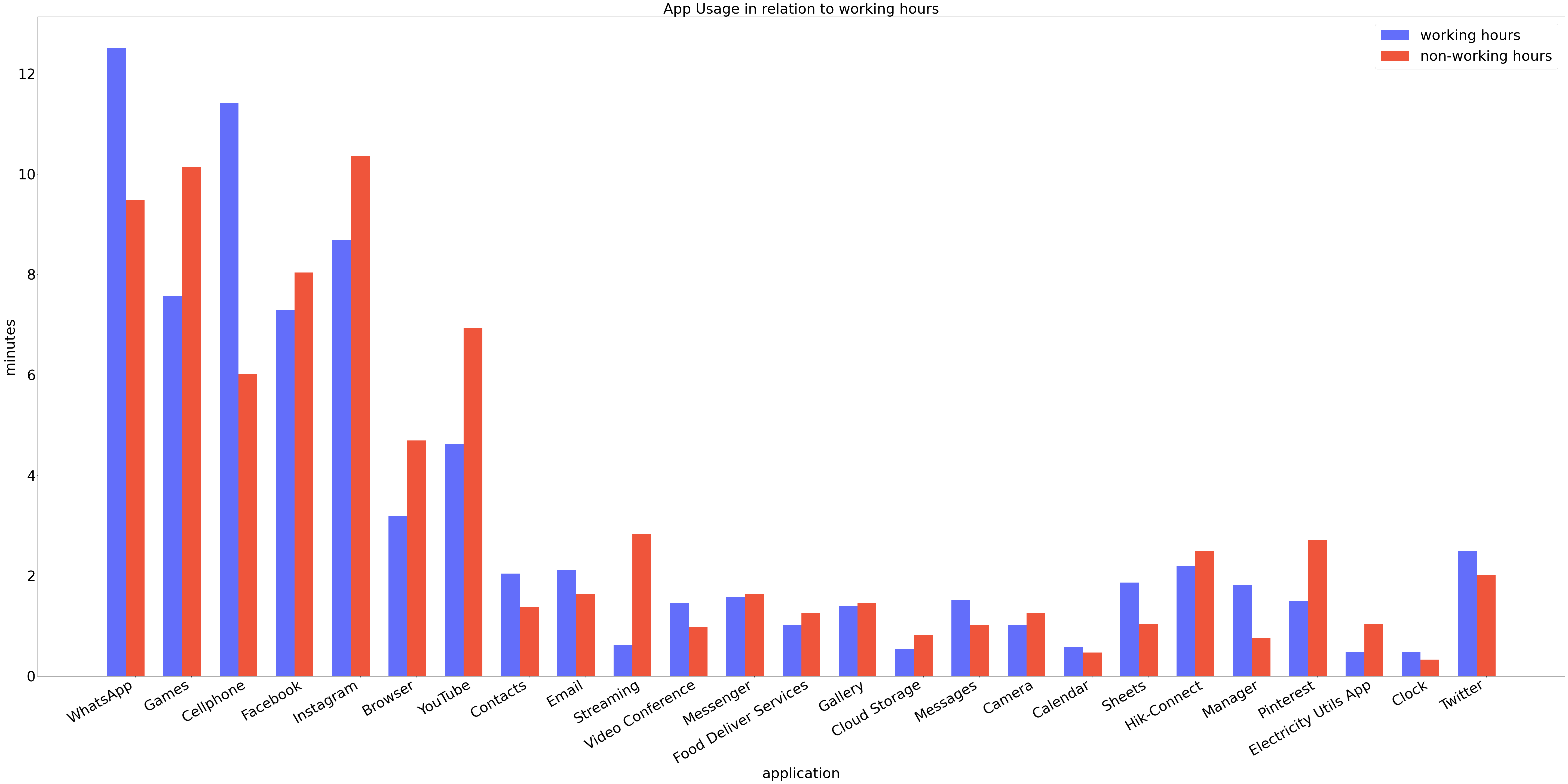}
\caption{Mean application usage during non-working and working hours.}
\label{fig:app_usage_work}
\end{figure*}

\subsection{Application purpose}
\label{sec:app_finality}

Like we have previously stated, an application can have different usage purposes depending on the user. This is also an aspect that we wanted to evaluate in this preliminary study, particularly the use of applications related to work and non-work activities. As we stated before, there are several applications that are highly tied to work activities, such as e-mail and office tools. However, there are other applications that even though their primary use is not for work, can be used as tools for such activities. On the other hand, most applications are not related to work activities and are mostly used for leisure. The use of the various types of applications, their relationship with work/non-work activities, and the time at which they were used during the day could offer some information about the users’ daily schedule.

To better evaluate these aspects, we wanted to further inspect how the most used applications relate to working and non-working periods. Since we did not have specific details about the participants’ work, or schedules, we considered the most common working schedule in Portugal. Normally in Portugal, the working schedule consists of 40 weekly working hours, with most people working eight daily hours during weekdays. Additionally, most people work from 8 to 18h with a 2-hour break for lunch. As such, this was the arrangement considered when dividing the data in working periods and non-working periods. Additionally, all weekends were considered as nonworking periods.

The mean usage time of each application, in minutes, divided by these two periods can be seen in Figure \ref{fig:app_usage_work}. In this figure we present the mean value for the total duration of the study, as we only aimed to evaluate the use of different types of applications in relation to working periods and nonworking periods, and not their evolution.

\begin{figure*}[ht]
\centering 
\includegraphics[width=7in]{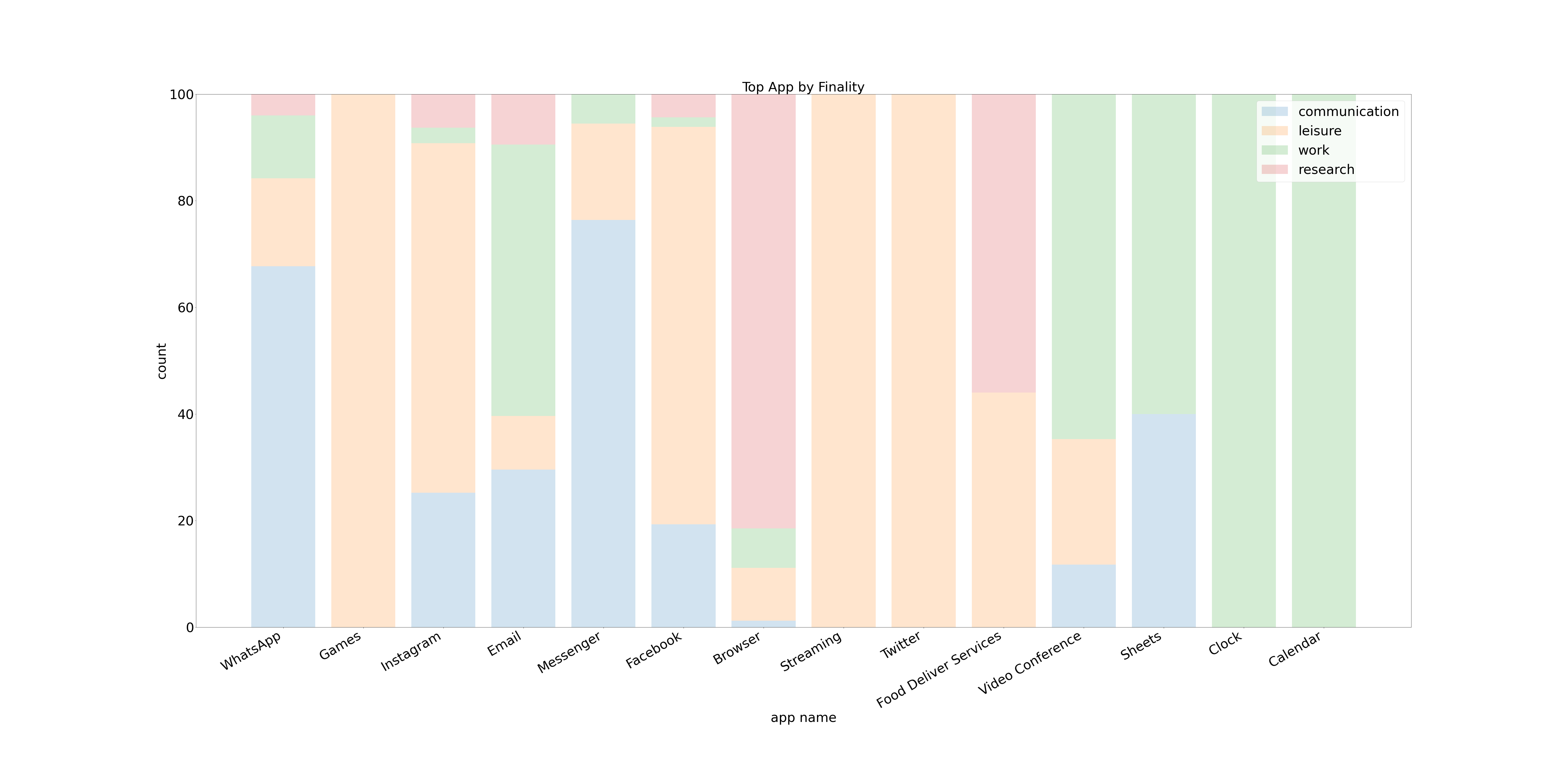}
\caption{Type of use in percentage for most used applications}
\label{fig:app_finality_percent}
\end{figure*}

We can see from the figure that applications used for interpersonal communication, like WhatsApp, Cell phone, Video Conferencing and E-mail tend to be more used during working periods. This was expected since most workflows, especially when working remotely, require that people use virtual means of communication with co-workers. However, we can also see significant usage of these applications outside normal working hours. This was also expected, as these applications are also used for personal communications outside work, like WhatsApp and the Cell phone. However, the high usage of Video Conference and Email applications outside normal working hours could point to a shift in the working schedule when working remotely.

Furthermore, as expected, apps related with leisure tended to be used more often outside of working hours. This was the case of Games, Streaming applications, YouTube, and social networks. However, as we can also see from Figure \ref{fig:app_usage_work}, there was also a considerable usage of Game applications and Social Network applications during working hours. This could point to the participants taking some time for small breaks within their working hours, which can happen more easily due to working from home. Additionally, when analysing the list of distinct games present in the Games category, we verified that many of those games seemed to target very young children. In Portugal, during the duration of this trial study, schools had adopted a remote lecture scheme as well, leading to many parents having to manage their work schedule along with their children’s activities. In fact, we believe that some of the usage of gaming applications was due to participants trying to entertain/distract their children for some time, as some studies have pointed out the rapid increase in the usage of smartphones and other similar devices by small children \cite{radesky2015mobile}. However, as previously stated, without more personal information from the participants we cannot draw firmer conclusions about this.

Additionally, as explained in section \ref{sec:active_data}, the participants were prompted to answer a daily questionnaire and select the purpose that better described their use of the five most used applications during each day. Since the users were only prompted with the most used applications for that particular day, and these applications vary between users and days, the list of these applications may differ from the previously presented 25 most used applications presented in Figure \ref{fig:app_usage_both_periods}. As such, only 14 of the total most used applications are presented in this data. The percentage value of each type of purpose per application can be seen in Figure \ref{fig:app_finality_percent}. Note, that Figure \ref{fig:app_finality_percent} scale carries no relation with the scales of Figures \ref{fig:app_usage_both_periods} and \ref{fig:app_usage_work}. That is, a small percentage in one type of purpose may not indicate a small mean usage time for that application and that particular purpose, and vice versa, since the user is only prompted to answer the purpose the of five most used applications in the day, and the total usage time can vary greatly from one day to another, or even from one user to another.

As can be seen in Figure \ref{fig:app_finality_percent}, some applications were exclusively or almost exclusively used for one purpose. For instance, Games, Twitter and Streaming applications are used exclusively for leisure activities, while others like the Clock or Calendar applications are exclusively used for work. Other applications, like Email, Sheets and Video Conference applications are mostly used for work.

By comparing the answers to this questionnaire with the data shown in Figure \ref{fig:app_usage_work} we can see that some of the applications used mostly for leisure, such as Games, Twitter and Facebook, were used almost indistinctly during work and non-work periods. This could mean that participants felt the need to take breaks, due to working from home and having more control over their schedules. Applications mostly used for communication, such as WhatsApp and Messenger, presented the same behaviour. However, for this type of applications, some users report using them for work-related activities, which could explain some of the usage during work periods.

On the other hand, applications that are mostly used for work such as the Clock, Calendar, Sheets, and Email, were used on both working and non-working periods. This could point to a difficulty in \textit{”disconnecting”} from work. Other studies have explored the challenges of working remotely, and pointed out the difficulty in self-managing time \cite{flores2019understanding}, which could lead to working more hours and/or later. Additionally, it is not uncommon for employees to contact workers outside working hours, especially in a remote working context, potentially giving some people the impression that they are always at work. In fact, in November 2021, Portugal deemed it illegal for employers or companies to call or message their workers outside normal working hours \cite{Portugal0:online}. However, more information about the participants’ working schedule is needed to draw conclusions, and this is something which will be addressed in future work.

\subsection{Correlation with Covid feeds of Information}
\label{sec:correlation_covid}

\begin{figure*}[ht]
\centering 
\includegraphics[width=6.5in]{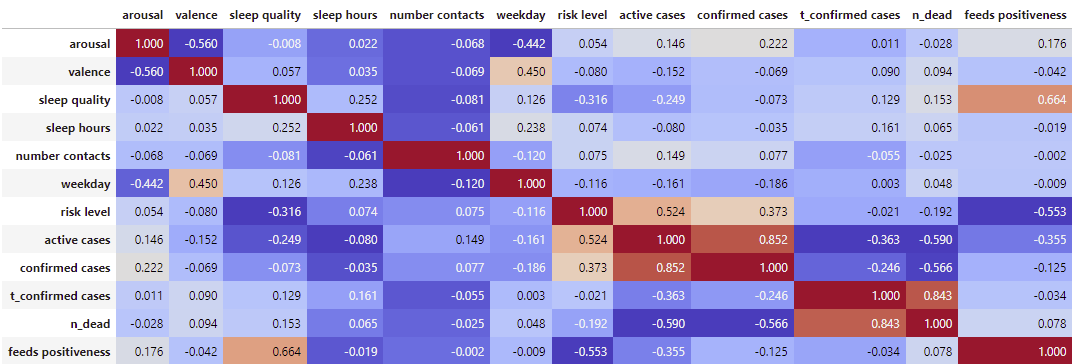}
\caption{Time-lagged Cross-Correlation Matrix of collected data and Covid related numbers and events (3 days shift).}
\label{fig:correlation_matrix}
\end{figure*}

The way information is disseminated, and the particular information that is shared with the public in a situation such as the one lived during the Covid-19 pandemic, can directly influence peoples’ lives, their psychological state and even their physiological states. As preliminary analysis, we intended to explore the correlation between public information feeds/significant events and the collected data. In these preliminary results we compared the data obtained from the mobile questionnaires, namely the sleep questionnaire, the Emotional Questionnaire (\gls{SAM} scale) and the transport and proximity questionnaires, with the most used metrics by the media to describe the pandemic situation. Furthermore, we compared this data with the general public opinion in relation to the most relevant Covid-related events, presented in Table \ref{Table:events_covid}, and with several Covid-related numbers that were disclosed by the competent Portuguese authorities on a daily basis. In addition, the correlation of these variables with weekdays was also considered.

\begin{figure}[!t]
\centering 
\includegraphics[width=3in]{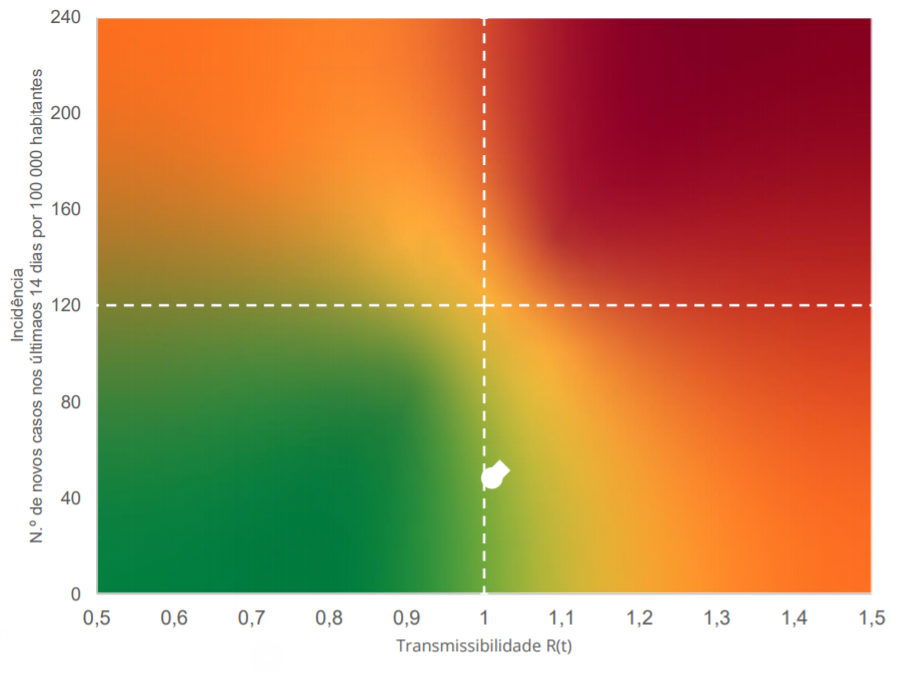}
\caption{Example of Covid Risk Matrix, Extracted from a \gls{DGS} report.}
\label{fig:risk_matrix_example}
\end{figure}

Considering the data collected from the various questionnaires, it is possible to derive five different features. From the emotional questionnaire, it was possible to obtain the daily value of valence and arousal for each participant, which can represent the emotional state of a person. The data collected from the sleep questionnaire, included the daily sleep quality and number of slept hours. And lastly, both the transport questionnaire and proximity questionnaire data were fused to obtain a daily total number of possible contacts by user. For each variable, a time series was generated, considering the mean value by day and by user.

During the pandemic, the Portuguese National Health System maintained a dashboard where daily reports concerning the evolution of pandemic situation and numbers in Portugal were released. From those reports, it was possible to extract a dataset that is now publicly available \cite{dssgptco39:online}. We selected some features from that dataset that were the most used by the media, namely, the values for the number of active cases, new confirmed cases, total confirmed cases, and number of new death cases for each day. Additionally, a metric not directly present in the dataset but that was one of the most used by the media was the risk level, using a risk matrix that could be calculated from the incidence and the transmissibility of the virus (Rt value), with the top right quadrant/dark red representing the highest level and the bottom left/light green representing the lowest level. An example of this risk matrix can be seen in Figure \ref{fig:risk_matrix_example}. We calculated the risk level values from the incidence and transmissibility values available in the dataset, using the following formula:

$$
\begin{cases}
 & \text{2 } if \text{ incidence}>  120  \text{ AND transmissibility} > 1  \\ 
 & \text{1 } if \text{ incidence}> 120  \text{ OR transmissibility} > 1  \\ 
 & \text{0 } if \text{ incidence}\leq 120  \text{ AND transmissibility} \leq 1  \\ 
\end{cases}
$$

In addition, we made a compilation of the most relevant Covid-related events in Portugal between the months of January and May 2021, which can be seen in Table \ref{Table:events_covid}. In this table it is also possible to see the positiveness of each event, i.e., events marked as \textit{"+"}, were perceived as positive by the general public, while the ones marked \textit{"-"} were perceived as negative. We then proceeded to calculate the positiveness index of each date, by considering the +1 value for positive events and -1 for negative ones. The value for each day was then the sum of all of the values that preceded that day. This could be seen as an accumulation of events in people’s memory, and a relative shift in their perception of the state of the pandemic.

Furthermore, we are aware that the state of mind of people does not change immediately on receiving information, and that their reactions can be influenced by short-term history. Given this, we decided to evaluate the Time-lagged Cross-Correlation, using the Pearson coefficient, between the acquired data and the Covid-related metrics/events. In this particular case we wanted to evaluate the response of people to the current or short-lived feed of information about the pandemic. As such, the data acquired from people was stationary, and all related Covid features suffered a shift in time. We calculated the time-lagged Cross-Correlation with a shift of zero to four days, being that the best results were obtained for a shift of three days. Due to the overflow of new daily information, we believe that it is not important to explore a time-lag greater than four days.

Considering the data presented in Figure \ref{fig:correlation_matrix}, we can see that, apart from the arousal and valence values, none of the collected data has correlations between them. Furthermore, we can see that the valence and arousal values were strongly and negatively correlated, which indicates that the participants mostly reported emotions that fall within the pleasure and anger sectors of the 2D emotion diagram \cite{2013yazdani}. That is, when the participants felt more positive, they also felt less emotionally energetic, which indicates relaxation, serenity, or contentment. On the other hand, when they felt more energetic, they also felt less positive, which indicates feelings of annoyance and anger. In addition, both valence and arousal had a moderate correlation with the weekdays, namely a negative one for arousal and a positive one for valence. This was to be expect since most people tend to fell less energetic as the week progresses, and normally the weekend is associated with leisure time, when people tend to feel more positive. However, this information also serves as validation for the collected data, since it correlates with the findings of other studies \cite{helliwell2014weekends}.

Considering the correlation between the acquired data, and the time-lagged Covid news/events metrics, the only strong correlation was between the reported sleep quality and the total positiveness of past events with a shift of 3 days. This indicates that people reported that they slept better when the past events related with Covid were more positive. Furthermore, we can see from the coefficients that the number of slept hours and the sleep quality did not have any significant correlation, which indicates that this increase would be mostly due to the participants average perception of the Covid-related events.

Additionally, there was also a moderated negative correlation between the sleep quality and the zone of the risk matrix with 3 days shift. As previously stated, the risk level of the risk matrix was one of the most used metrics by the National Health Organizations and the media. As such, this correlation could indicate that the information broadcasted by the media could affect the way people rated their sleep quality. Note there is also a strong correlation between the risk level and the total positiveness, which could explain this correlation. 

On the other hand, the other metrics used by the media, such as the number of new cases and the number of deaths had no significant correlation with the collected data. As previously stated, the risk level matrix was given considerable importance by the Portuguese media. This was due to the fact that the risk level is a more stable metric than the number of cases and number of deaths. Moreover, this metric was directly linked to the deconfinement measures and, thus, had direct implications on peoples’ lives, which can explain the differences in terms of the existence of a correlation.

Sleep quality was the only metric from the collected data that presented a significant correlation. This could indicate that self-reported sleep quality might be a good indicator for how well people dealt with the reported Covid situation and events. Furthermore, during the trial, the self-reported emotional data indicates that the participants mostly experienced feelings from the anger and the pleasure sectors of the emotional diagram. However, as previously noted, a larger sample size is needed to draw stronger conclusions about the interplay of these factors. Additionally, it would be interesting to compare the collected data with data collected in a period during which the lockdown and the pandemic are not active. Both of these concerns will be addressed in future studies.

\section{Conclusion}
\label{sec:conclusions}

In this paper we presented an innovative solution to monitor humans’ physical and emotional states, through a smartphone and smartwatch application, that can be used in the context of the SarsCov-2 pandemic and in possible future pandemics. Furthermore, we present the outcomes of a project that involved both online questionnaires through the Qualtrics\textsuperscript{TM} platform, as well as a period of data collection using the developed application. The collection of data through online questionnaires included 333 participants, in three waves corresponding to August, October and November 2020. Alternatively, the collection of data using the mobile and smartwatch applications were performed with 14 participants for three and a half months, from February to May 2021. As a result, we were able to collect several types of data from several different periods of the pandemic. As far as we known, our study is the first to use a mobile application to collect passive data, as a complement to a larger online study.

Although there are some studies that explore different aspects of peoples’ lives such as activity and emotional states, as well as some works that explore the effects of the SARS-CoV-2 on people, our application is the first that is able to explore multiple aspects simultaneously, as well as to collect data directly related with risk factors concerning the ongoing pandemic. Additionally, and most importantly, our system is able to provide tailored feedback to each user regarding all of the acquired data.

The collected data allowed us to identify some behavioural changes in the subjects during the pandemic, including changes in terms of the used mobile applications. Additionally, possible correlations between our data and Covid-19 related metrics and events largely broadcasted in the media were identified and analysed. 

Future work will explore two main avenues. On one hand, we intend to further enhance the application, to make it more flexible and adaptable to other purposes and environments, and to further explore its human-in-the-loop features. On other hand, we intend to perform trials with higher numbers of subjects, not only to strengthen the support for the preliminary conclusions drawn in this study, but also to study ways to incentivise and reward user participation.


\ifCLASSOPTIONcompsoc
  \section*{Acknowledgments}
\else
  \section*{Acknowledgment}
\fi

Part of the work was also financed by the National Funds provided by the Portuguese Foundation for Science and Technology (FCT) through the project “ResiliScence 4 COVID-19: Social Sensing \& Intelligence for Forecasting Human Response in Future COVID-19 Scenarios, towards Social Systems Resilience” (Research 4 COVID-19 - Project n. 439) and through the CISUC R\&D unit UID/CEC/00326/2020 and programmatic funding code UIDP/00326/2020. The authors would also like to thank Dr. Samuel Domingos for his important feedback about the variables to measure through the Vitoria app and the design of tailored feedback. José Marcelo Fernandes wishes to acknowledge the Portuguese funding institution FCT - Foundation for Science and Technology for supporting his research under the Ph.D. grants SFRH/BD/147371/2019.  




%


%

\bibliographystyle{ieeetr}
\bibliography{sample-base}

\end{document}